\documentclass[reviewl,3p,twocolumn]{elsarticle}

\usepackage{graphicx}
\usepackage{epstopdf}

\graphicspath{{./figures/}}

\usepackage{caption}
\usepackage{subcaption}
\usepackage{epsfig}

\usepackage{fixltx2e}
\usepackage{array}
\usepackage{amsmath}
\usepackage[shortcuts]{extdash}

\usepackage{amssymb}
\usepackage[pdftex, colorlinks]{hyperref}
\hypersetup{colorlinks=true, breaklinks=true}

\usepackage[switch,modulo]{lineno}

\biboptions{compress}

\journal{Astroparticle physics}

\begin{document}

\sloppy
\begin{frontmatter}



\title{The SPHERE-2 detector for observation of extensive air showers in~1~PeV~--~1~EeV energy range}
\author[inst1]{R.A.~Antonov}
\author[inst1]{E.A.~Bonvech}
\author[inst1]{D.V.~Chernov\corref{cor1}}
\ead{chr@dec1.sinp.msu.ru}
\author[inst1]{T.A.~Dzhatdoev}
\author[inst2,inst3]{M.~Finger~Jr.}
\author[inst2,inst3]{M.~Finger}
\author[inst1,inst4]{D.A.~Podgrudkov\corref{cor1}}
\ead{d.a.podgrudkov@physics.msu.ru}
\cortext[cor1]{Corresponding author}
\author[inst1]{T.M.~Roganova}
\author[inst1]{A.V.~Shirokov}
\author[inst1,inst4]{I.A.~Vaiman}
\address[inst1]{Skobeltsyn Institute for Nuclear Physics of Lomonosov Moscow State University, Moscow, Russian Federation}
\address[inst2]{Charles University, Faculty of Mathematics and Physics, Prague, Czech Republic}
\address[inst3]{Joint Institute for Nuclear Research, Dubna, Russian Federation}
\address[inst4]{Physics Department of Lomonosov Moscow State University, Moscow, Russian Federation}

\begin{abstract}
The SPHERE-2 balloon-borne detector designed for extensive air shower (EAS) observations using EAS optical Vavilov-Cherenkov radiation (``Cherenkov light''), reflected from the snow-covered surface of Lake Baikal is described. We briefly discuss the concept behind the reflected Cherenkov light method, characterize the conditions at the experimental site and overview the construction of the tethered balloon used to lift the SPHERE-2 telescope above the surface. This paper is mainly dedicated to a detailed technical description of the detector, including its optical system, sensitive elements, electronics, and data acquisition system (DAQ). The results of some laboratory and field tests of the optical system are presented.
\end{abstract}
\begin{keyword}
detector \sep PMT \sep electronics \sep Vavilov-Cherenkov radiation \sep extensive air showers
\end{keyword}
\end{frontmatter}

\section{Introduction} \label{sec:intro}

In the energy range of $E>1$~PeV, primary cosmic rays (CR) are routinely studied via observation of cascades initiated by them in the atmosphere --- the so-called extensive air showers (EAS). Precise evaluation of EAS parameters is required in order to study the primary CR composition. Unfortunately, at $E>100$~PeV dense ground-level EAS arrays such as KASCADE \cite{Antoni2005}, EAS-TOP \cite{Aglietta2004} and CASA-BLANCA \cite{Fowler2001} give low statistic results due to the very low flux of primary CR of such energies. At such energies in view of the limited experimental budget one is forced to either drop the quality of data or to come up with other methods of EAS registration.

Observation of reflected optical Vavilov-Cherenkov radiation (``Cherenkov light'') of EAS is a promising technique for CR studies in the 1~PeV--10~EeV energy range~\cite{ANTONOV201924,Antonov:2015xta}. This approach allows to measure the shape of the lateral distribution function (LDF) near the axis ($R<50$~m)~\cite{Antonov:2015aqa}. This LDF region is believed to be the most informative part of the LDF in terms of composition sensitivity~\cite{ANTONOV201924, sphere2009FIAN-eng}. Additionally, parameters of the Cherenkov light LDF are typically much less dependent on the high energy hadron interaction model than the parameters of the muon component of EAS.

The SPHERE experiment is the first successful attempt of registration and reconstruction of a considerable number of EAS with the reflected Cherenkov light (RCL) method~\cite{Antonov:2015xta}. These observations were made with the SPHERE-2 detector at Lake Baikal using a tethered balloon (``BAPA'' --- a transliterated abbreviation from Russian ``Baikal Tethered Balloon''). 

\begin{figure}[t]
\includegraphics[width=0.475\textwidth]{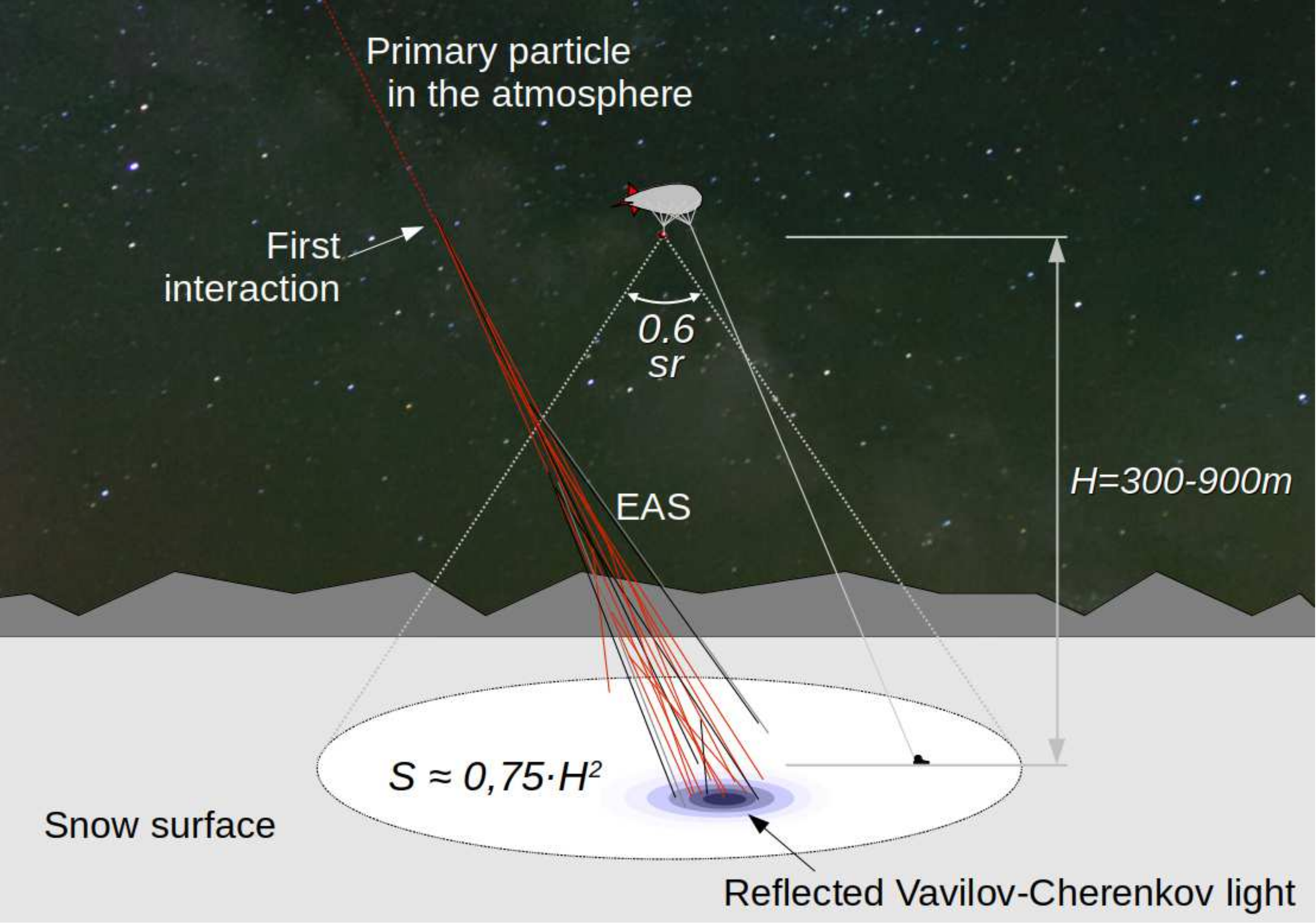}
\caption{Concept of reflected Cherenkov light observation in the SPHERE experiment.}
\label{experiment_scheme_new}
\end{figure}

The concept of the SPHERE experiment is illustrated in Fig.~\ref{experiment_scheme_new}. The SPHERE-2 detector observes EAS Cherenkov light by overlooking a part of the Lake Baikal surface with an area $\approx(3/4)\cdot H^{2}$. The direction of the primary particle is reconstructed using the timing information and its energy is estimated using full Cherenkov light flux and shape of the LDF \cite{Antonov:2015aqa}. The statistical uncertainty of the energy determination is 10--20~\% depending on the energy and the observation conditions. Finally the fraction of the CR low mass component could be identified using the shape of the LDF \cite{ANTONOV201924,Antonov:2015xta,Antonov:2015aqa}.

This paper is mostly devoted to the technical description of the SPHERE-2 detector. Also the observation conditions and the construction of the BAPA balloon are described. The applicability of the RCL approach to EAS observations is discussed. It was demonstrated that both all-nuclei spectrum (subsection 6.1 of~\cite{Antonov:2015xta} and \cite{sphere2013JP-results}) and composition~\cite{Antonov:2015aqa} could be successfully reconstructed with the reflected Cherenkov light approach. A more detailed analysis with an improved account of systematic effects is underway and the results of this analysis will be published elsewhere.

\begin{figure}[t]
\center{\includegraphics[width=0.475\textwidth]{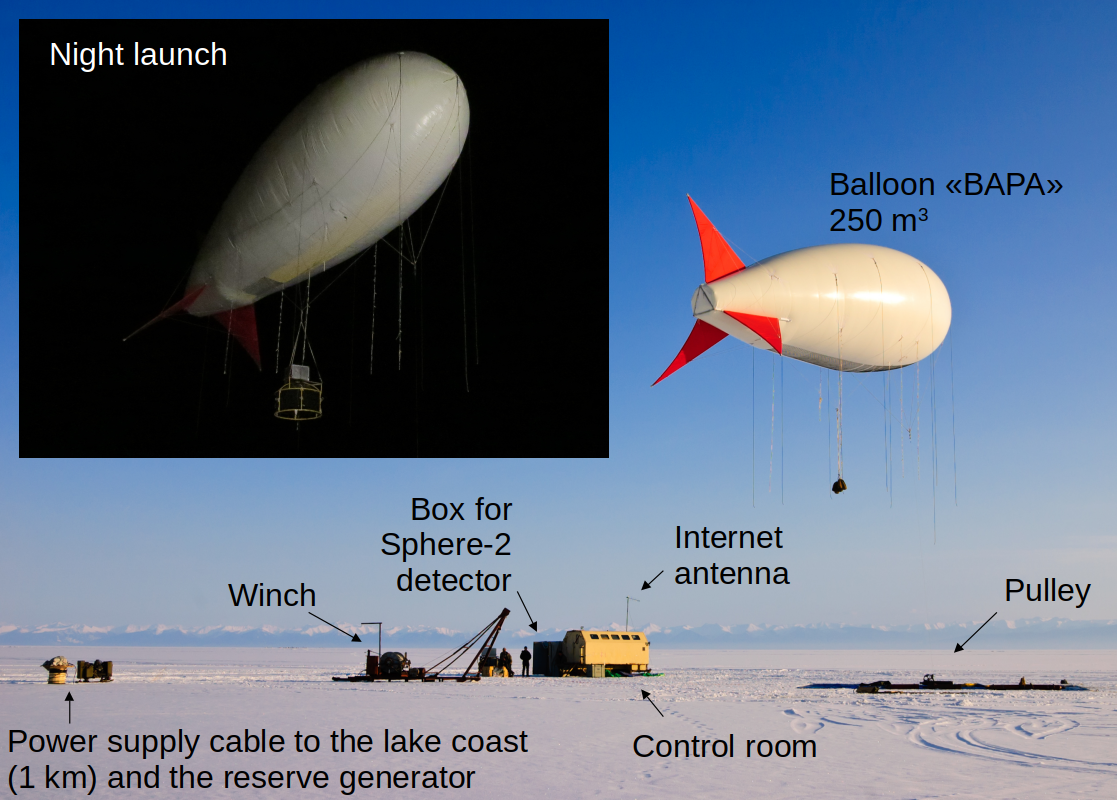}}
\caption{``BAPA'' tethered balloon together with the ground-based launch pad equipment. The balloon is shown during a daylight test flight with an equivalent load to adjust the attack angle. Inset (top left) shows the SPHERE-2 detector suspended under the BAPA balloon during a night flight.}
\label{balloon}
\end{figure}

\section{Conditions at the experimental site}\label{sec:conditions}

All measurements with the SPHERE-2 detector were carried out in the Baikal National Park. This area is characterized by a low level of air and light pollution
. The launch pad for the tethered balloon was located on the ice of Lake Baikal at a distance of about 1 km from the shore in a place with coordinates $N$~$51^{\circ}47'49''$, $E$~$104^{\circ}23'19''$, 455~m above sea level. This place is located a few kilometers from the Baikal neutrino telescope \cite{GVD2018}. The site of the SPHERE experiment is accessible by the Circum-Baikal railway, which is currently rarely used; its presence did not influence the measurements.

Good transparency of the atmosphere and the absence of moonlight are critical factors for the successful operation of the SPHERE experiment. The transparency was checked visually ensuring absence of clouds and visibility of the Milky Way during the observation runs. A thick enough ice cover (at least 50~cm) is also needed in order to support the heavy load of vessels filled with helium (at least 3 t) and other equipment necessary to launch the balloon (see Section 3 for more details). These conditions are fulfilled at Lake Baikal at the end of February and in March.

Finally, the proposed method of EAS registration is also sensitive to the properties of the snow surface, a detailed discussion of which was performed in \cite{ANTONOV201924} and it was shown that the uncertainties associated with the snow surface do not impair the results of Cherenkov light measurements and reconstruction of EAS parameters. During the measurements, the stability of the snow reflection coefficient was periodically monitored with a luxmeter.

Taking into account the lunar cycle and ice conditions it is possible to carry out one or two measurement sessions annually, each being about 10 days long. On average about half of the nights during these sessions are suitable for measurements.

\begin{figure}[t]
\includegraphics[width=0.48\textwidth]{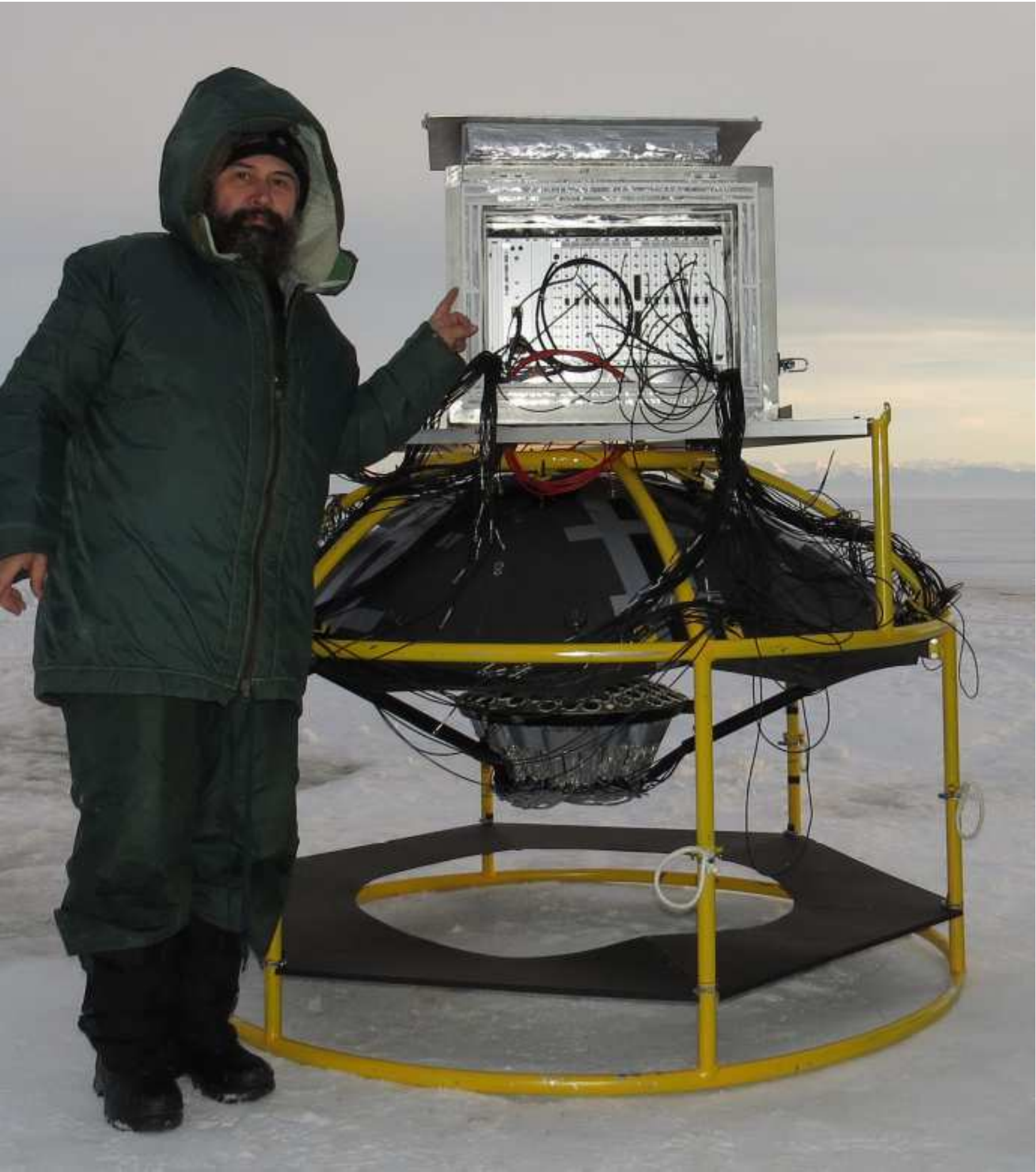}
\caption{SPHERE-2 telescope on the surface of Lake Baikal (March 2013).}
\label{detector_installation}
\end{figure}

\section{The BAPA balloon} \label{sec:aerostat}

The BAPA tethered balloon (see  Fig.~\ref{balloon}) was developed by the Avgur-RosAeroSystems company~\cite{rosaerosystems} specifically for the needs of the SPHERE experiment. BAPA is a variable-volume balloon that has a wide range of working pressures. The dimensions of the balloon are as follows: length --- 15.57~m, diameter --- 5.75~m; its initial/maximum volume is 225/250~m$^3$. The contraction system consisted of 133 rubber cables that provided 200~Pa and 585~Pa overpressure values (i.e. the pressure above the external air pressure) at the minimum and maximum balloon volumes, respectively. For further control of the overpressure an automatic vent was installed which opened at the overpressure values of about 750~Pa and shut close at about 650~Pa. A differential pressure and temperature sensor was also installed inside the balloon volume (see description below). This allowed to select the optimal working altitude and to control the vertical speed during the initial ascent.

The 250~m$^3$ shell of the ``BAPA'' balloon was made from Lamcotec~\cite{lamcotec} nylon-based laminated gas-tight fabric. The shell was supplemented by an inverted Y-type three plane empennage for flight stabilization. According to the manufacturer the balloon had the maximum flight altitude of 1000~m with working temperatures at that altitude ranging from $-30^\circ$C up to $+20^\circ$C and the working wind speed under 20~m/s. The maximum payload was 80~kg with a 30~kg reserve at the 1~km altitude.

The balloon was held by a 3.1~mm diameter and 41.1~g/m density steel cable with the estimated maximal working load of 800~kg. The balloon's rigging slings were attached to the cable via a swivel to ensure free rotation of the balloon allowing  orientation along the wind. On the other end the cable went through a pulley fixed on the ice surface and to an electric winch. 

\begin{figure}[t]
\includegraphics[width=\linewidth]{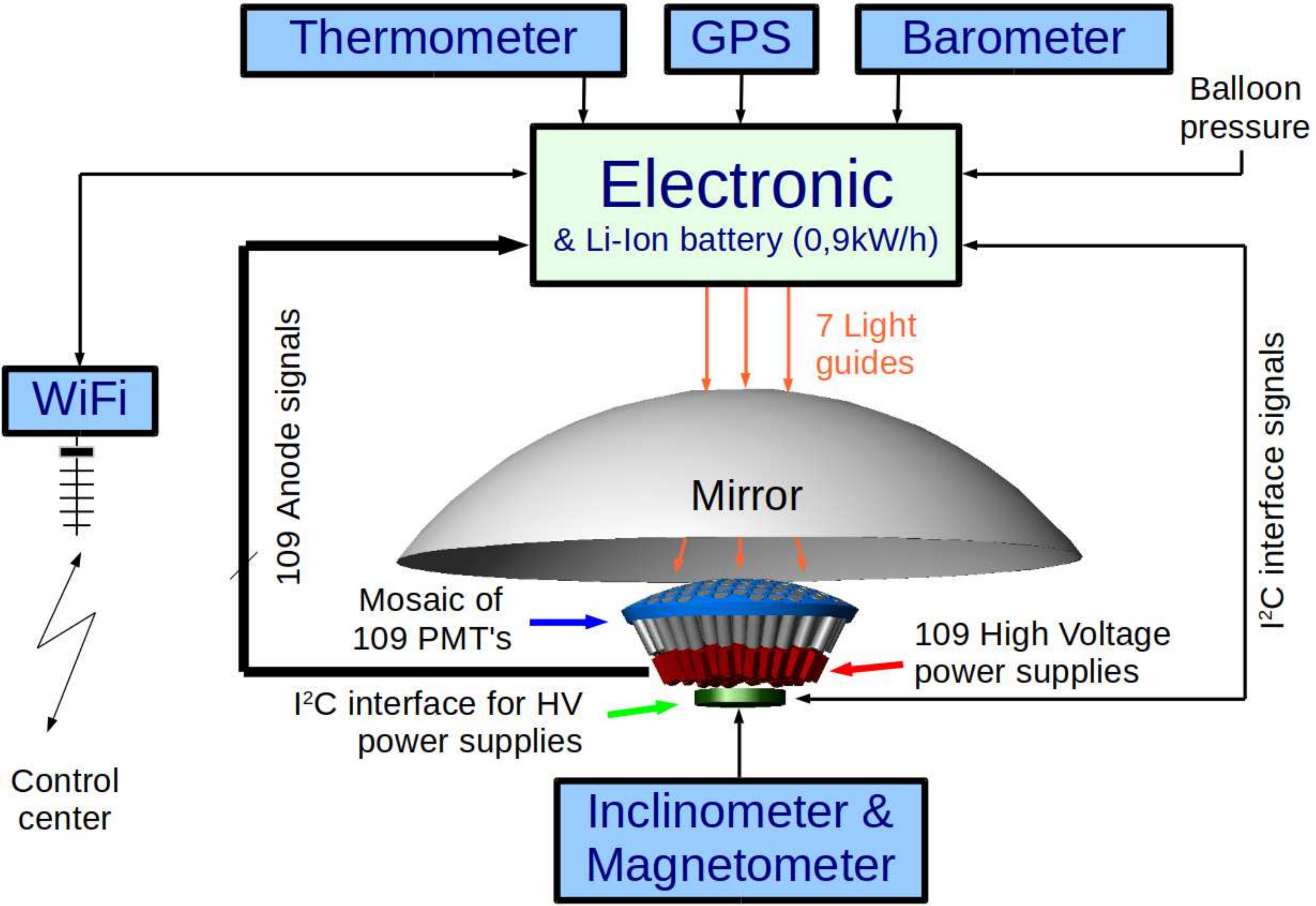}
\caption{Scheme of the SPHERE-2 detector.}
\label{electronics_modules}
\end{figure}

\section{The SPHERE-2 detector}\label{sec:detector}

The assembled SPHERE-2 telescope is shown in Fig.~\ref{detector_installation}. The principal scheme of this detector is presented in Fig.~\ref{electronics_modules}. It consisted of a spherical mirror, photomultiplier tube (PMT) mosaic, and a control block. Each PMT had its own signal cable. The PMT power supply modules were installed directly onto each PMT and were controlled through the I$^2$C interface~\cite{I2C_spec} individually using a commutator located on the mosaic.

\subsection{Optical system}

\begin{figure}[t]
\center{\includegraphics[width=0.45\textwidth]{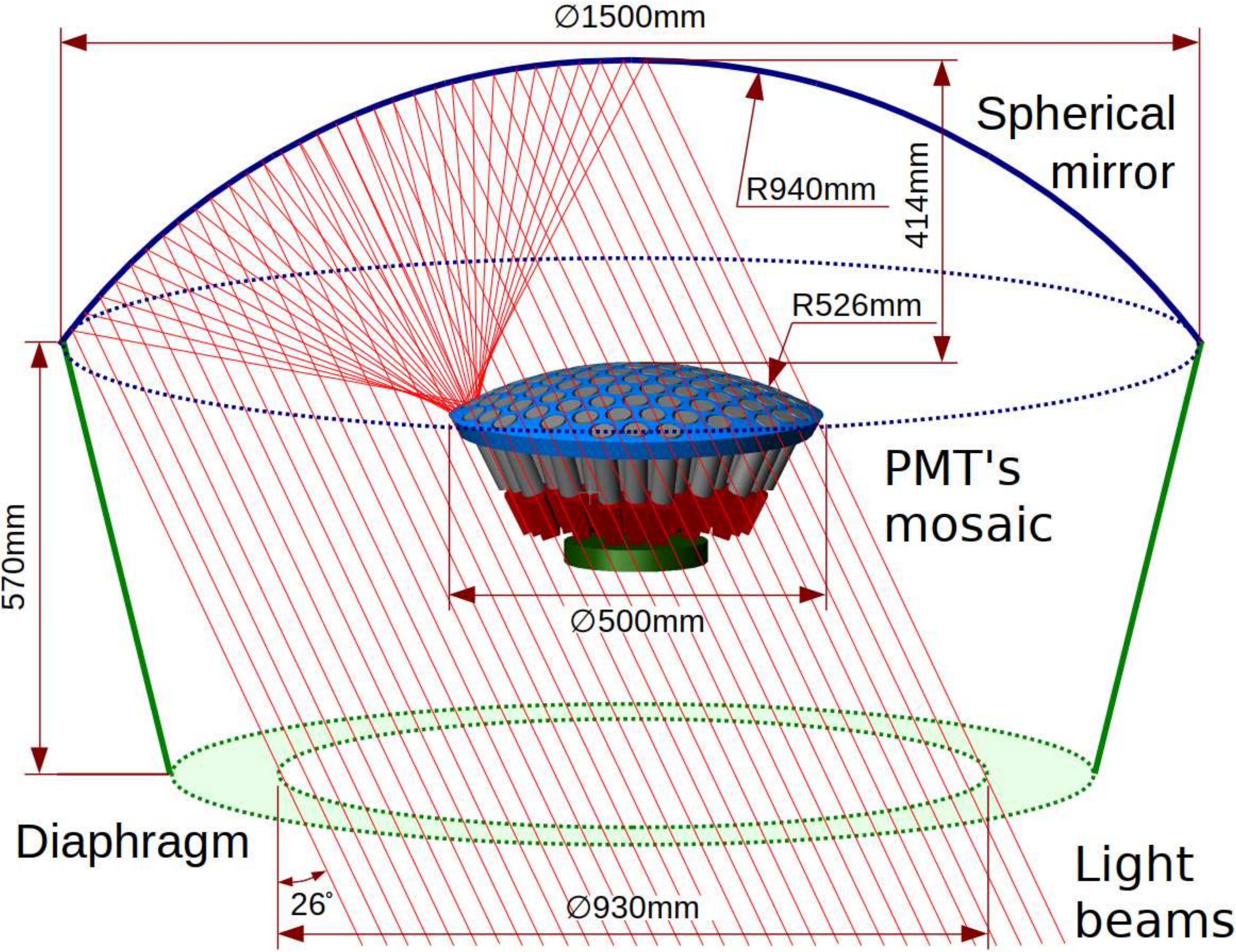}}
\caption{Optical scheme of the SPHERE-2 telescope.}
\label{optic}
\end{figure}

The SPHERE-2 telescope (see  Fig.~\ref{optic}) is based on the Schmidt optical system without a corrector plate. This simplified design suffers from spherical aberrations, but is suitable for our aims since the area of the focused light beam is comparable to the spatial resolution of the mosaic.

\begin{figure}[t]
\includegraphics*[width=\linewidth]{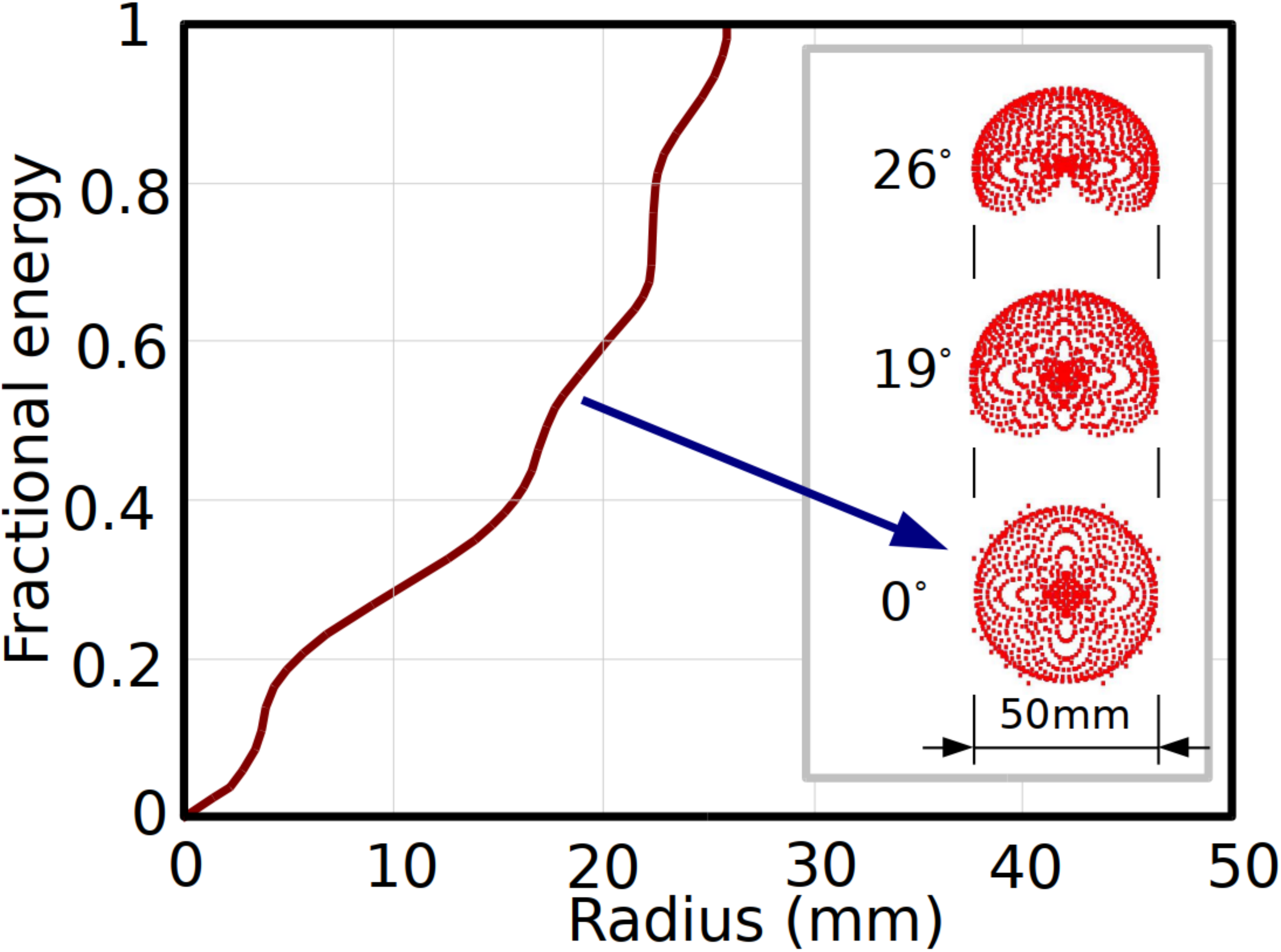}
\caption{PSF distribution function for the normal incidence. Inset (right) shows the image of the PSF for the normal incidence (bottom), as well as for the incidence angle of 19$^{\circ}$ (middle) and 26$^{\circ}$ (top).}
\label{optic_beams}
\end{figure}

The mirror of the SPHERE-2 telescope had a diameter $D$= 1.5~m, the curvature radius $R_{c}$= 0.94~m and consisted of seven segments. The segments were produced from multilayered carbon fabric, coated with a 120~nm reflective aluminium layer and a 40~nm transparent protective SiO$_2$ layer. The integral representation of the point spread function (PSF) of the mirror, namely, the fraction of the total energy falling within a certain radius (the PSF distribution function) is shown in Fig.~\ref{optic_beams} for a light beam parallel to the detector axis. The PSF was calculated with the publicly-available OSLO EDU software (version 6.6)~\cite{OSLO} and is presented in Fig.~\ref{optic_beams} (inset) for three values of the incidence angle. All these calculations were performed for parallel beams. The asymmetry of the spot at angles 19$^\circ$ and 26$^\circ$ is due to a shadow from the mosaic. The total diameter of the PSF is about 50~mm, what is comparable to the distance between the centers of adjacent PMTs. Therefore, emission from any source on the snow surface inside the field-of-view (FOV) of the SPHERE-2 telescope would result in a non-negligible signal, irrespectively of the location of this source.

Optical properties of the mirror were tested using a light source based on an ultra-bright light-emitting diode (LED) APRL-20W-EPA-3040-PW installed in the center of a 100x100~mm screen behind a 5~mm diaphragm with a diffuser. The light source was installed on a test bench positioned on the optical axis of the apparatus near the mirror curvature center. The PSFs of all segments were similar, about 15 mm in diameter (this broadening of the reflection images is due to the non-ideal texture of the reflecting surface). The focal distances of all segments were found to be the same. Finally, the reflection images of the segments did not reveal any sign of mechanical deformation. 

\begin{figure}[t]
\includegraphics[width=\linewidth]{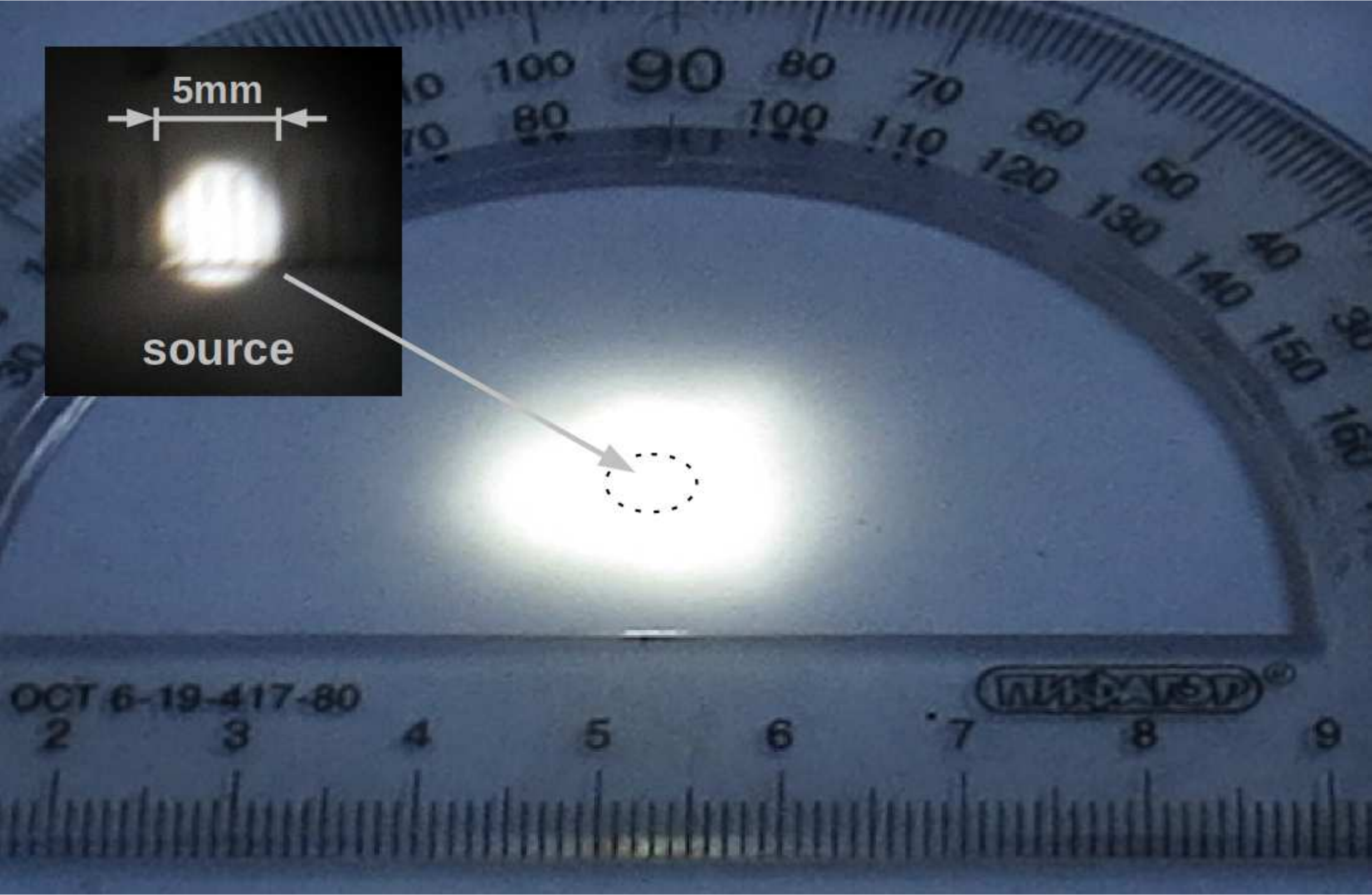}
\caption{Image of the LED light source formed with the assembled 7-segment mirror.}
\label{mirror_adjusting}
\end{figure}

Before the start of observations, all seven segments were assembled and once more tested using the same light source. The resulting image revealed a single 20~mm light spot (see Fig.~\ref{mirror_adjusting}). This mirror check procedure was performed outdoors with an air temperature close to the one recorded during test flights.

After the mirror was assembled, a 0.93~m-diameter  diaphragm was installed 100~mm above the mirror curvature center and the PMT mosaic was installed 420~mm below the mirror ($\sim$520~mm above the mirror curvature center and $\sim$50~mm above its focal point). The mosaic itself had a shape of a truncated sphere segment with the 526~mm curvature, 250~mm outer radius and 300~mm total height (see Fig.~\ref{optic}).

\subsection{PMT mosaic}

The mosaic consisted of 108 FEU-84-3~\cite{FEU84} PMTs and one Hamamatsu R3886~\cite{R3886} PMT. FEU-84-3 PMT has a 25~mm diameter multi-alkali (Sb-K-Na-Cs) photocathode with modulator and fine mesh type 12 stage dynode system. It is sensitive in the 300--800~nm wavelength region with a maximum in sensitivity around 420--550~nm. Typical peak quantum efficiency of FEU-84-3 PMT is around 18\%~\cite{Antonov2016}. The Hamamatsu R3886 PMT installed in the center of the mosaic had a 34~mm diameter bialkali (Sb-K-Cs) photocathode and a circular cage 10 stage dynode system. It is sensitive in the 300-650~nm wavelength region with a maximum of sensitivity about 25\% around 420~nm~\cite{Antonov2016}.

\begin{figure}[t]
\includegraphics[width=\linewidth]{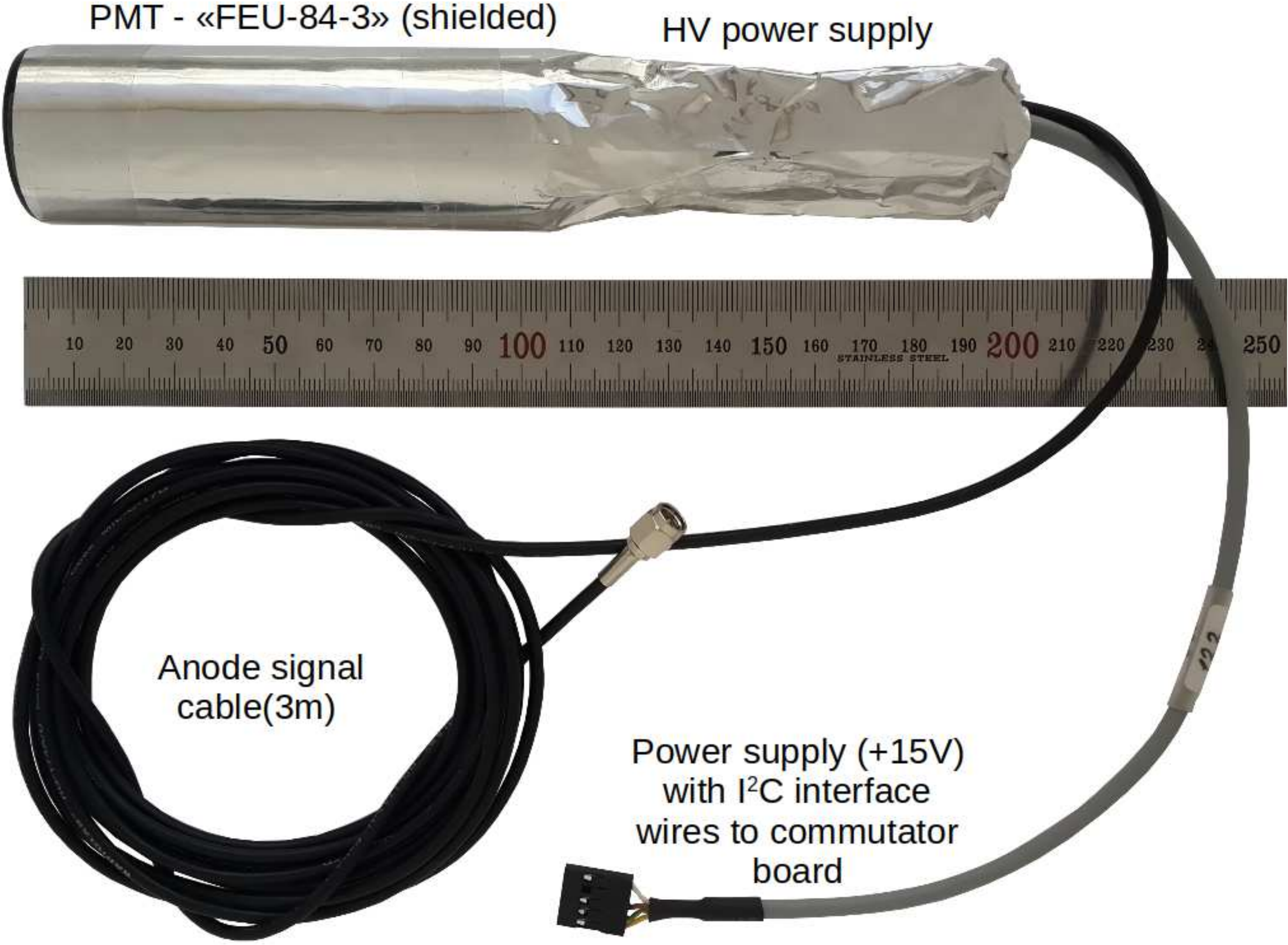}
\caption{Optical module with a PMT FEU-84-3 and high voltage power supply.}
\label{feu_84_3}
\end{figure}

Each PMT had a high voltage power source (HVPS) installed on it. A PMT with a power source formed an optical module (OM) (see Fig.~\ref{feu_84_3}). The OMs were tightly packed in the mosaic, therefore each OM was covered by an aluminium foil to reduce PMT crosstalks. For the OM powering and control a 30~cm four-wire shielded cable was used; two of the wires were utilized for the 15~V power input and other two wires for transferring commands and telemetry data via the I$^2$C interface. OMs were connected in parallel pairs (using odd and even addresses in a pair) to a commutator board, located directly on the mosaic. The analog output signals from all PMTs were transferred to readout electronics in the control block via 3~m RG-174A/U coaxial cables with 50~Ohm wave resistance. The PMT mosaic was shielded from wind, but was located outside the thermostatic container. The mosaic had a higher temperature than the outside air because of the heat from the high-voltage power sources.

\begin{figure}[t]
\includegraphics[width=\linewidth]{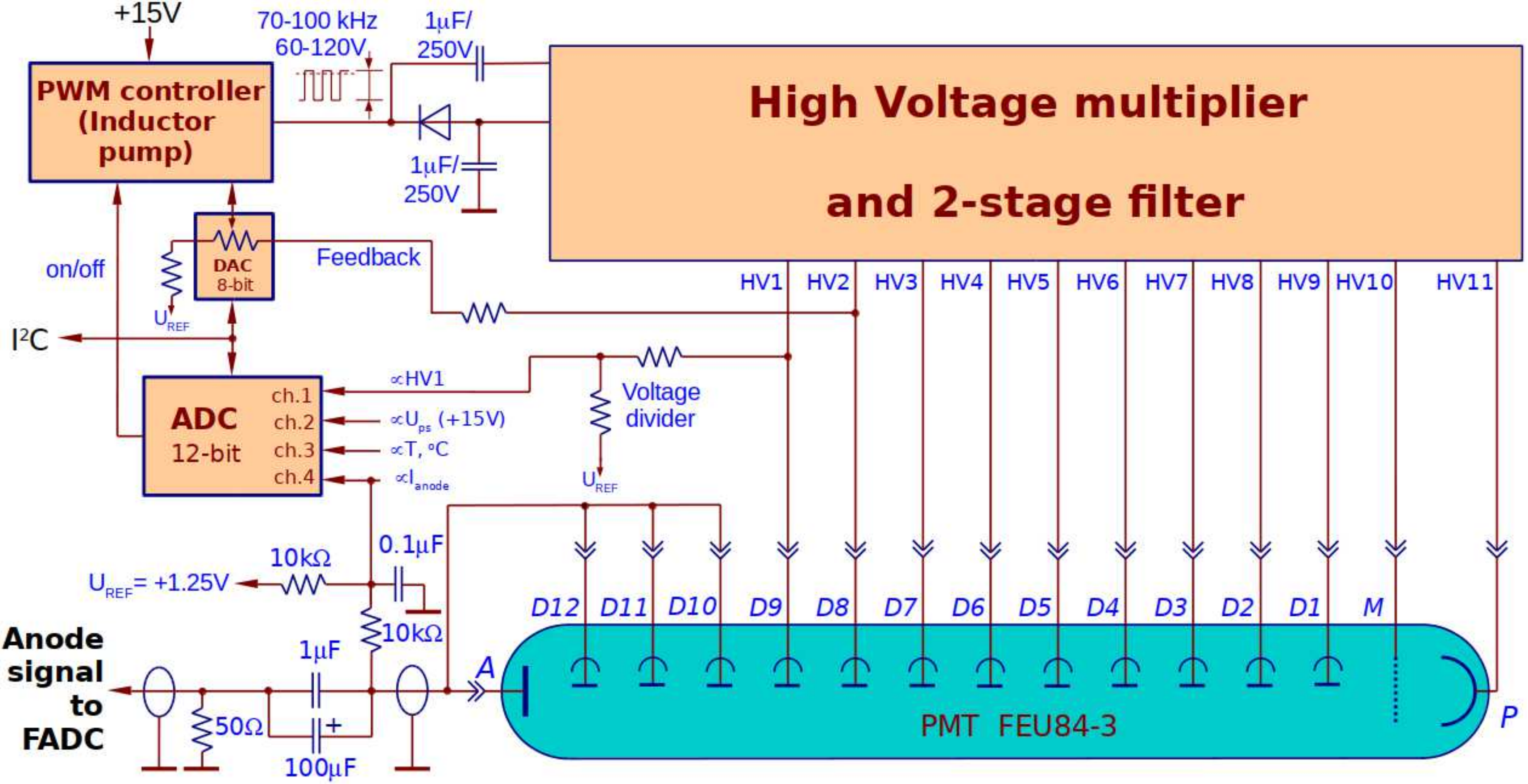}
\caption{Scheme of high voltage power supply for PMT.}
\label{HVPS}
\end{figure}

\subsection{High voltage power sources}

HVPS was a compact ($62\times25\times25$~mm) device with 11 output pins for FEU-84-3 connection. The scheme of the HVPS is shown in Fig.~\ref{HVPS}. The HVPS consisted of two boards: a voltage multiplier and a HVPS control board. The voltage multiplier is of Cockroft-Walton scheme with two RC circuits on every dynode. These RC circuits allow to filter out low frequency voltage variations and high frequency noise, thus providing 2~mV stability of dynode voltage. The multipliers were constructed using 0.47~$\mu$F 240~V high voltage capacitors (GRM43DR72E474KW01 \cite{murata}) and SM4005PL \cite{MCC} diodes with a maximum reverse voltage of 600~V. The filters were constructed using capacitors of the same lineup but with different voltages corresponding to those of the dynode voltages. All multiplier output pins were directly connected to the PMT pins except for the HV10 output which was connected to the PMT modulator for effective collection of photoelectrons on the first dynode. The voltage multiplier board was coated with thick layers of silicon sealant for electric isolation and leaks and discharge suppression.

The HVPS control board housed an inductive pump MAX1847~\cite{MAX1847} for high voltage pulse generation and a 8-bit Analog Devices AD5245BRJ50~\cite{AD5245BRJ50} digital to analog converter (DAC) chip for high voltage control in the 800--1500~V~range. This voltage was set according to the value recorded in the output register via the I$^2$C interface. The board controlled all voltages relative to the HV2 output value.

Another 4-channel 12-bit Analog Devices AD7994BRU~\cite{AD7994BRU} slow ADC chip was used to control the HV1 output voltage, +15~V input voltage, PMT average anode current with the precision of 0.1~$\mu$A and the power source board temperature with a B57621C0474J062 NTC thermistor~\cite{EPCOS}. The digital output ALERT/BUSY of the AD7994BRU chip was employed to turn the inductive pump on and off. After the +15V power was turned on, the high voltage remained turned off. High voltage could only be activated with a command via the I$^2$C interface.

The voltages on all other (HV2-HV11) outputs were not measured during the experimental runs but were estimated using the HV1 voltage. The dependence of HV2-HV11 voltages from HV1 voltage was studied on a batch of random HVPSs.

The  power consumption of the HVPSs was less than 90~mW at a time-averaged PMT anode current of 100~$\mu$A. Under dark conditions (no light on the PMT photocathode) HVPS power consumption droped to 35~mW.

\subsection{Mosaic commutator}
\label{sec:commutator}
\begin{figure}[t]
\includegraphics[width=\linewidth]{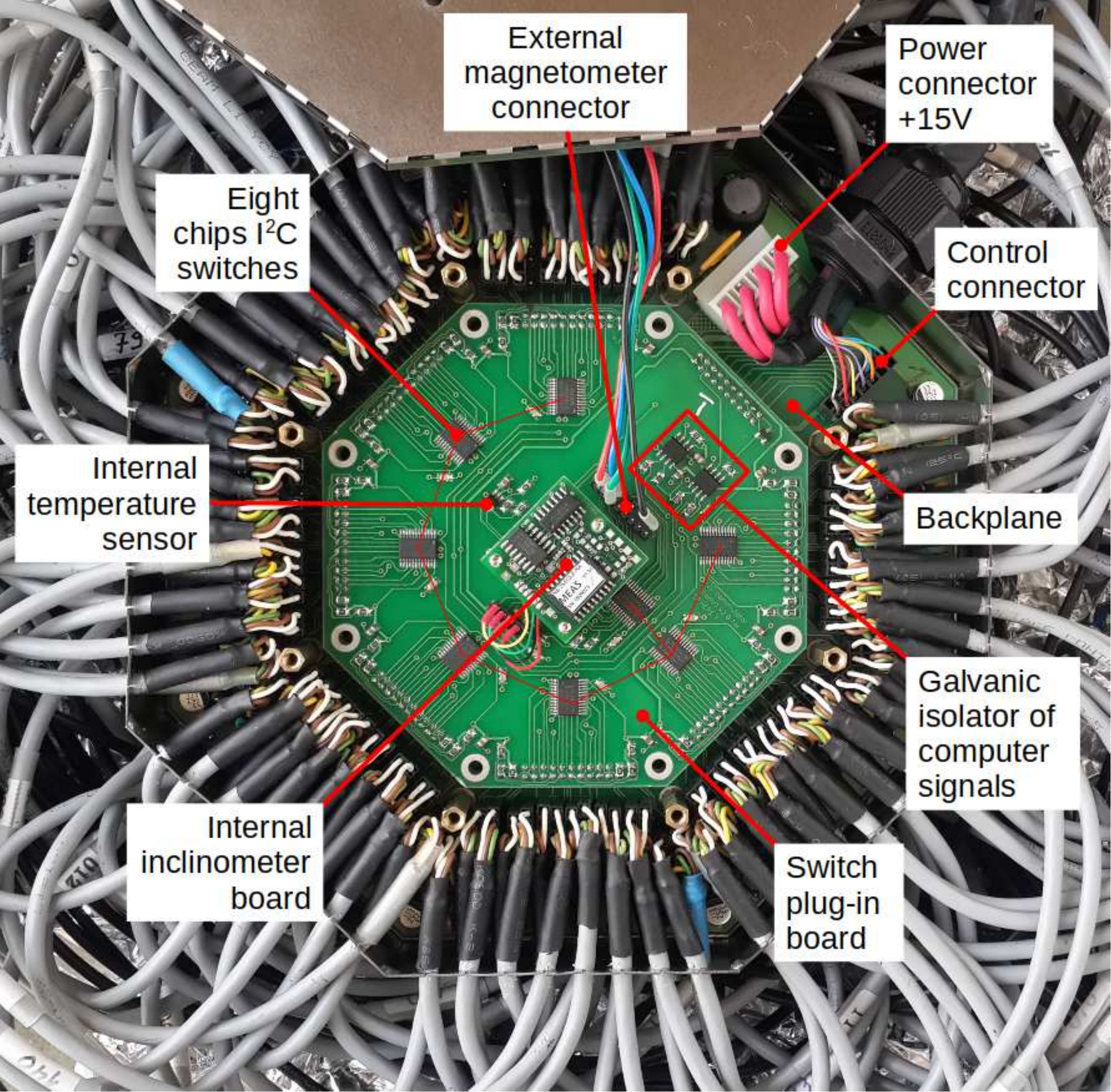}
\caption{The 64-channel $I^2C$ commutator board mounted on the PMT mosaic.}
\label{i2c_switch_board}
\end{figure}

The mosaic commutator was installed below the OMs (see Fig.~\ref{i2c_switch_board} for a general view of the commutator board) and consisted of two boards: the backplane with all the passive elements (filters, fuses and sockets for HVPS power and control connection) and a plug-in main board housing all chips, sensors, +3.3~V power stabilization, etc. This design was used to facilitate the replacement of failed active components without a full disassembly of the mosaic.

\begin{figure}[t]
\includegraphics[width=\linewidth]{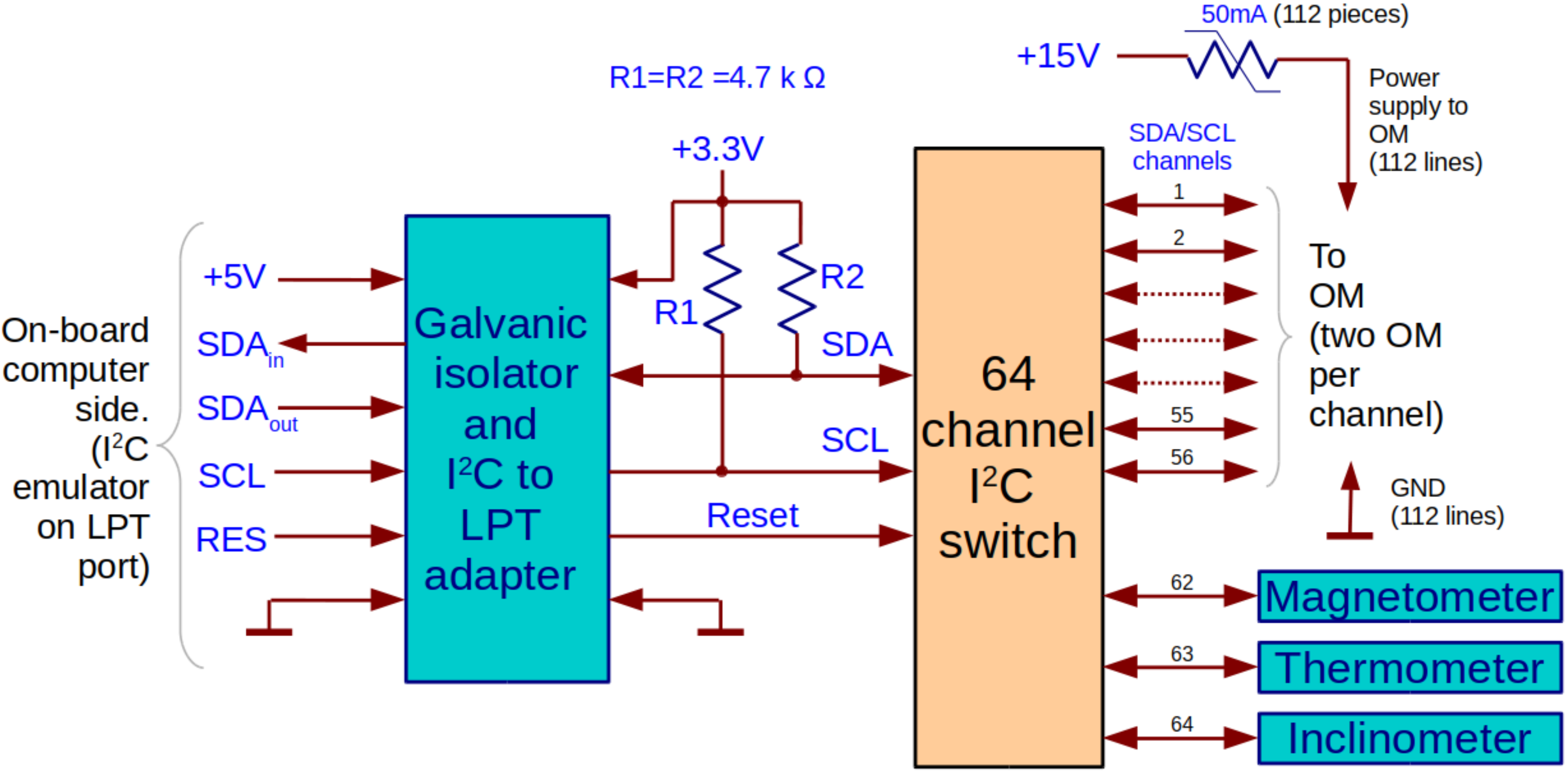}
\caption{Functional scheme of the 64-channel $I^2C$ commutator board.  
}
\label{i2c_switch}
\end{figure}

A principal scheme of the commutator is shown in Fig.~\ref{i2c_switch}. The commutator served for on-board computer access to each of the OMs via the I$^2$C interface. Since the on-board computer lacked the native I$^2$C support the interface was emulated using LPT port signal lines. Two Analog Device ADuM1251 chips~\cite{ADuM1251} were used to translate the commands, one for SCL and SDA signals and another for the independent RESET signal that switched the commutator to the initial state (immediately after power-on).

The core of the  commutator consisted of eight NXP Semiconductor PCA9547PW chips~\cite{PCA9547}, each with eight I$^2$C channels (64 channels in total). Channels 1--56 were used for OM commutation. Since OMs with odd and even addresses can be connected to the I$^2$C interface in parallel the total of 112 OMs could have been used. The OMs were powered through the commutator board. Each OM was connected to a +15~V power supply bus through a resettable fuse MF-R005~\cite{MF-R005} with a 50~mA maximum current.

Channels 57--62 were reserved for various sensors. The Analog Devices AD7415 sensor~\cite{AD7415} registered the temperature on the mosaic. The PMT mosaic commutator had a magnetometer and inclinometer (see Fig.~\ref{electronics_modules}) to control the orientation of the mosaic (rotation and tilt, respectively). Since the SPHERE-2 detector was freely suspended under the ``BAPA'' balloon, the inclination of the mosaic should be taken into account in order to reconstruct the geometrical parameters of the shower.

A Honeywell HMC6352 magnetometer~\cite{HMC6352} (not in production now) was installed 18~cm below the commutator board to control the mosaic orientation in the horizontal plane with a $\pm$2.5$^\circ$ precision. A 2-axis NS-25/DQL2-IXA inclinometer~\cite{inclinometer} was installed directly on the commutator main board and allowed to measure the inclination of the detector relative to the horizontal plane with a 0.1$^\circ$ precision (0.3$^\circ$ over the entire temperature range). During the observation runs the detector inclination was about 3$^\circ$ in calm conditions (the detector suspension system was not very accurate) and up to 18$^\circ$ in windy conditions.

\subsection{Control block}
\label{control block}

\begin{figure}[t]
\includegraphics[width=\linewidth]{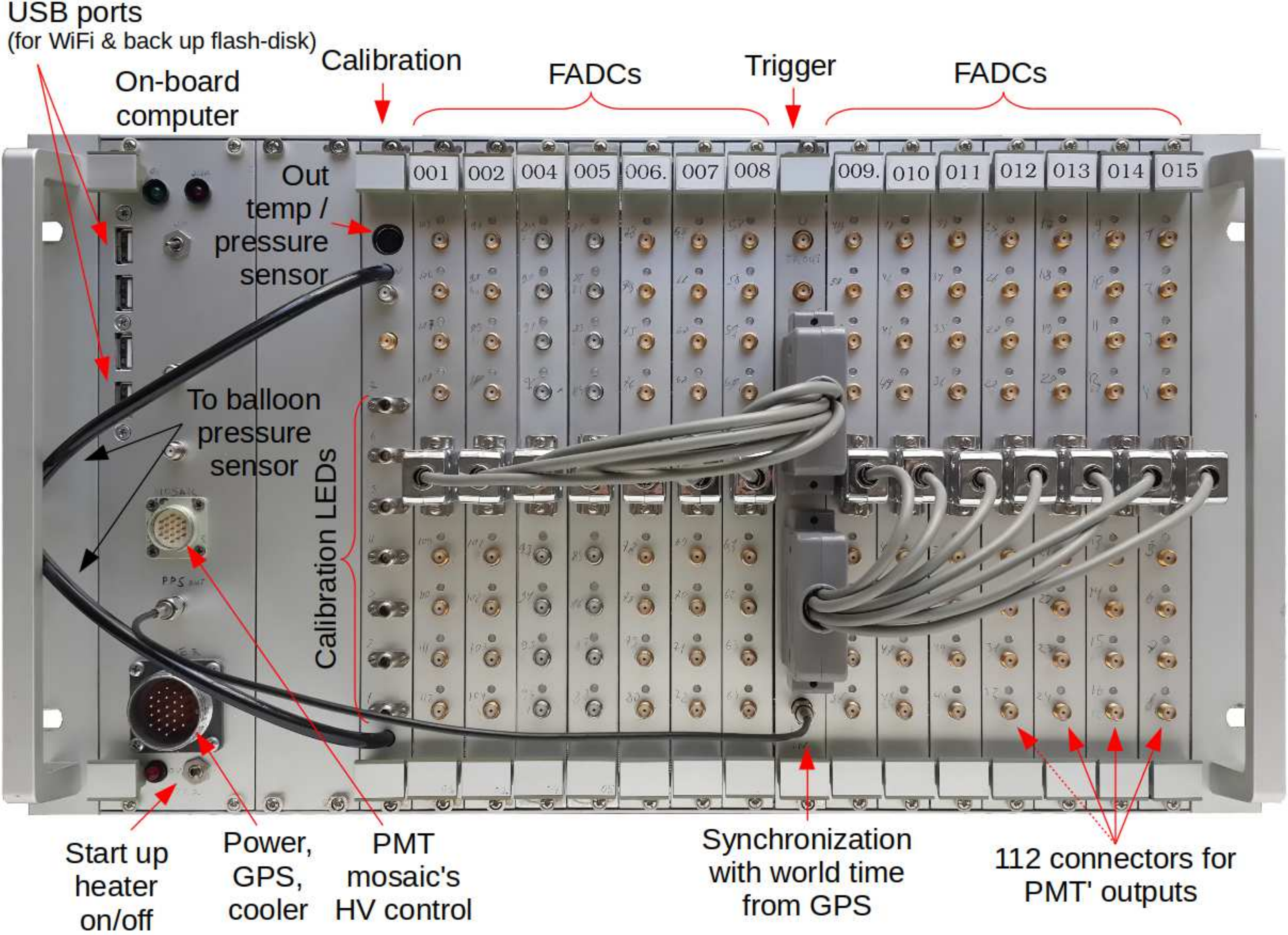}
\caption{Front panel of the electronics.}
\label{front_panel_of_electronics}
\end{figure}

The electronics were placed in a thermostable container above the optical part of the SPHERE-2 detector. All electronics were mounted in a 240~mm deep 19 inch crate with 21 slots of 6U high modules. A passive cross-ISA board PCA-6120~\cite{Advantech} with 20 slots was used for electronic board connections. The crate housed fourteen 8-channel FADC boards, the trigger board, LED calibration board (that also contained external pressure and temperature sensors) and an on-board computer~(see Fig.~\ref{front_panel_of_electronics}). The control block had two temperature sensors to measure the temperature outside and inside the block. In case of electronics overheating a cooling system was switched on. Another sensor was used to measure the temperature inside the balloon shell.

\subsubsection{Measuring channels}

\begin{figure}[t]
\center{\includegraphics[width=\linewidth]{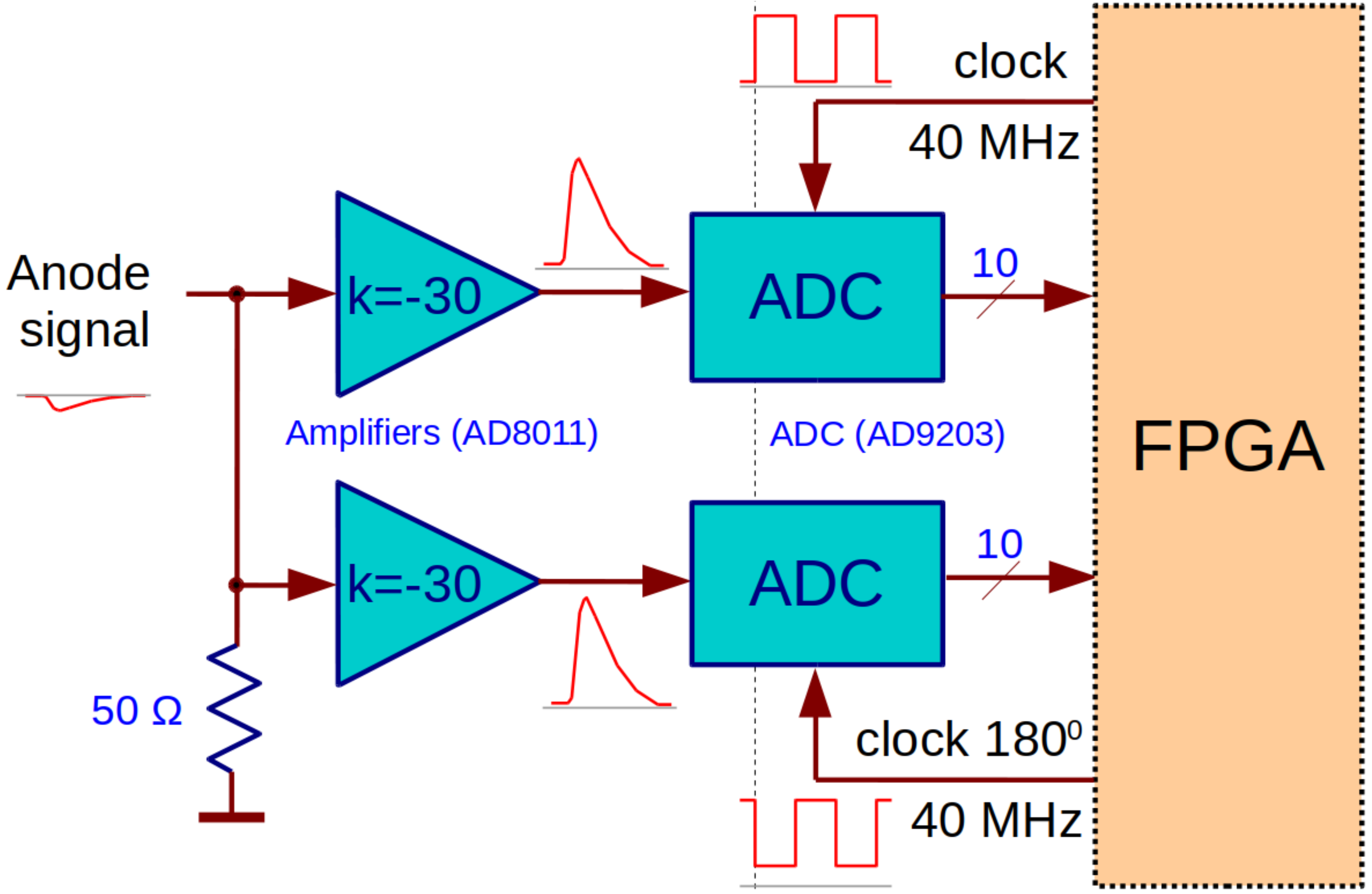}}
\caption{Functional diagram of a single channel FADC.}
\label{channel_scheme}
\end{figure}

In Fig.~\ref{channel_scheme} the logical scheme of the SPHERE\=/2 detector measuring channel is shown. The channel consisted of two 10-bit Analog Devices AD9203ARU ADCs~\cite{AD9203ARU} with 40~MHz sampling frequency and two AD8011 operational amplifiers~\cite{AD8011} with a multiplication factor of $-30.0\pm0.3$.The anode current signal from the OM was forked to these two amplifiers where it was inverted and amplified. Then the signal was fed into two ADCs which sent the digitized signal values to the FPGA chip every 25~ns. The synchronization signals were sent to the ADCs with a 180$^{\circ}$ shift between them, therefore the anode signal was digitized every 12.5~ns.

\begin{figure}[t]
\center{\includegraphics[width=\linewidth]{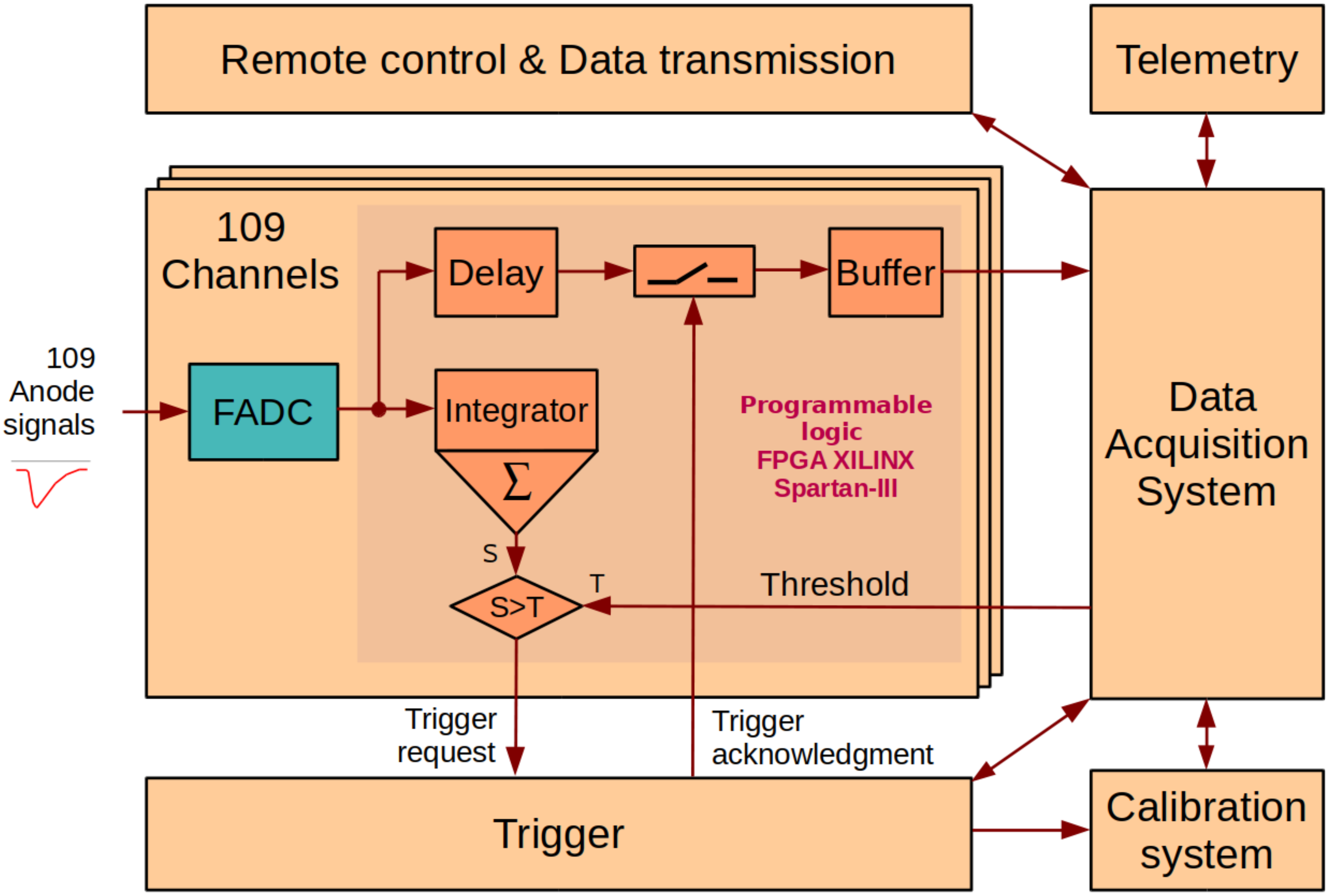}}
\caption{Data acquisition system (DAQ) scheme of the SPHERE-2 detector.}
\label{data_acquisition_electronics}
\end{figure}

The XILINX FPGA~\cite{Xilinx} chip (configuration bitstream chip) realized an algorithm of digital data flow handling. A sketch of this procedure is shown in Fig.~\ref{data_acquisition_electronics}. First, the incoming signals were forked into 6.4~$\mu$s delay lines. Then the signal from one of the two resulting branches was sent to an integrator that calculated the integral signal $A_{int}$ over the last 100~ns (the other branch was left out of consideration at this stage). This integral signal was then compared to a threshold $A_{thr}$ that was set individually for every channel. If $A_{int}>A_{thr}$ a discriminator signal was sent to the trigger board.

When the trigger board produced a ``trigger'' signal the information collected by the ADCs during the last 12.8~$\mu$s was copied to a specially-designed buffer. The use of the delay lines and the buffer allowed to record the signal not only around and after the trigger time, but also prior to this moment. The buffer is capable to store four events. The time needed to read one event from the buffer and to write this information to a solid-state drive (SSD) was approximately 0.25~s. Since every ``trigger'' event was accompanied by a calibration event, the maximal event reading frequency was about 2~Hz.

The discriminator signals were sent to the trigger continuously. If the trigger system produced a trigger signal while the buffer was full the DAQ system would've set the trigger flag to 1 and recorded 12.8~$\mu$s of data from the channel right after the buffer was freed. This resulted in detector ``dead time''. Such a situation is not typical for the SPHERE-2 detector operation conditions, but still occurs sometimes due to OM crosstalks or high illumination level.

\begin{figure}[t]
\center{\includegraphics[width=0.45\textwidth]{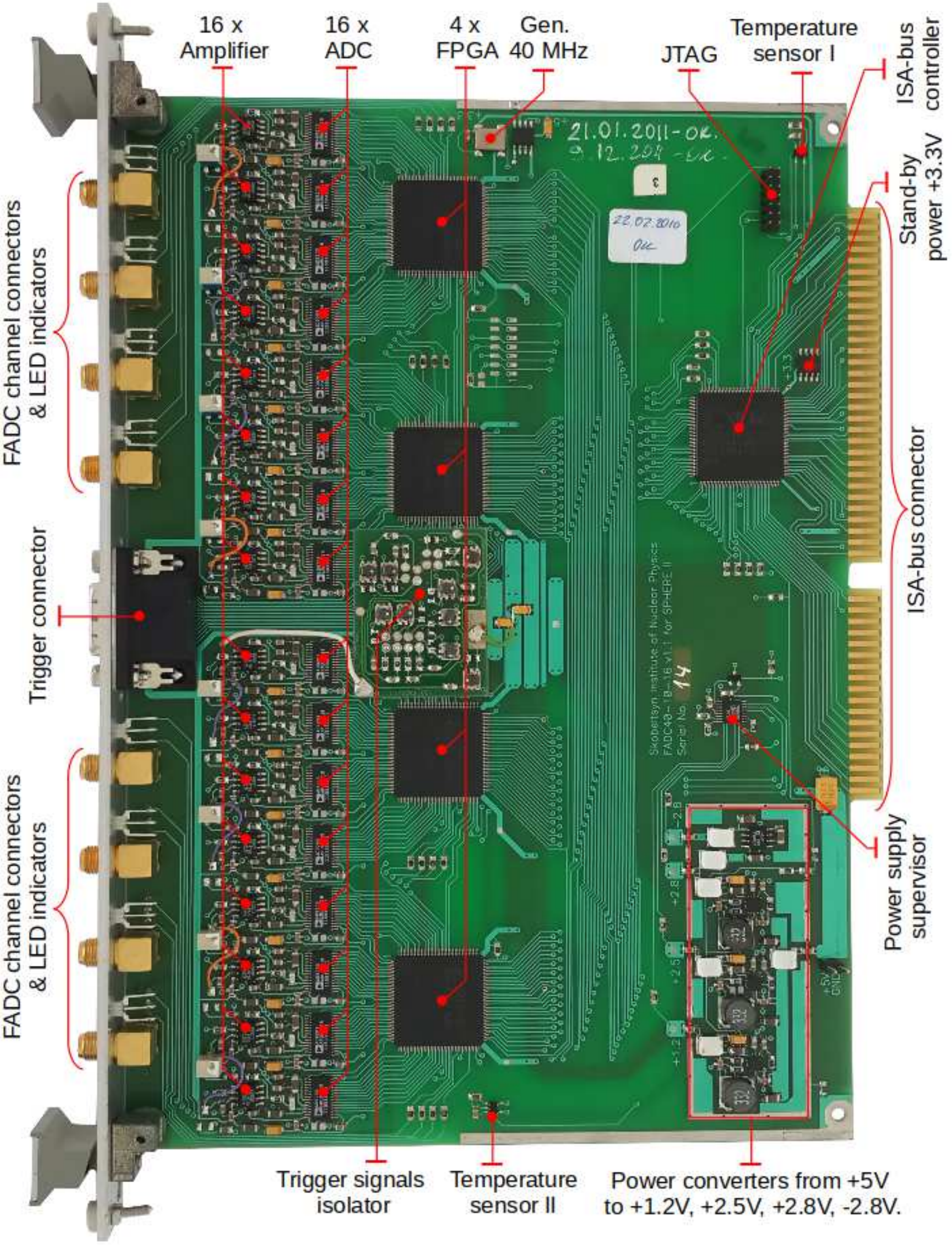}}
\caption{8-channel FADC board.}
\label{channels_board}
\end{figure}

A measuring channel board shown in Fig.~\ref{channels_board} had 8 input SMA connectors for analog signals. The board consisted of four XILINX Spartan3 XC3S200-4TQ144I~\cite{XILINX_Spartan} FPGA chips (one chip per four ADCs or per two channels). An XILINX CoolRunner XCR3128XL-10TQ144I~\cite{XCR3128XL} CPLD chip was used as an ISA bus client controller and program loader for FPGA.

Each flash ADC channel board had an individual secondary power supply. ISA bus voltage +5~V was converted to four output voltages +2.8~V, -2.8~V, +2.5~V and +1.2~V. For positive voltages a MAX1556~\cite{MAX1556} step-down DC-DC converter chip with an up to 97\% efficiency and maximal current of 1.2A was used. For negative voltage a switched-capacitor voltage converter ADM8660~\cite{ADM8660} was used that inverted +2.8~V to -2.8~V. The total power consumption of the board in full operation mode was below 2~W. Sleep mode was used when the FADC board was booted but awaited for the measurements to start. In this mode all converters were turned off and power consumption was around 10~mW. Only the CPLD chip worked as a supervisor and continuously awaited for commands. Also there was a possibility to measure temperature conditions with two sensors and power supply voltages with the additional AD7994BRU ADC chips~\cite{AD7994BRU}.

\begin{figure}[t]
\center{\includegraphics[width=0.45\textwidth]{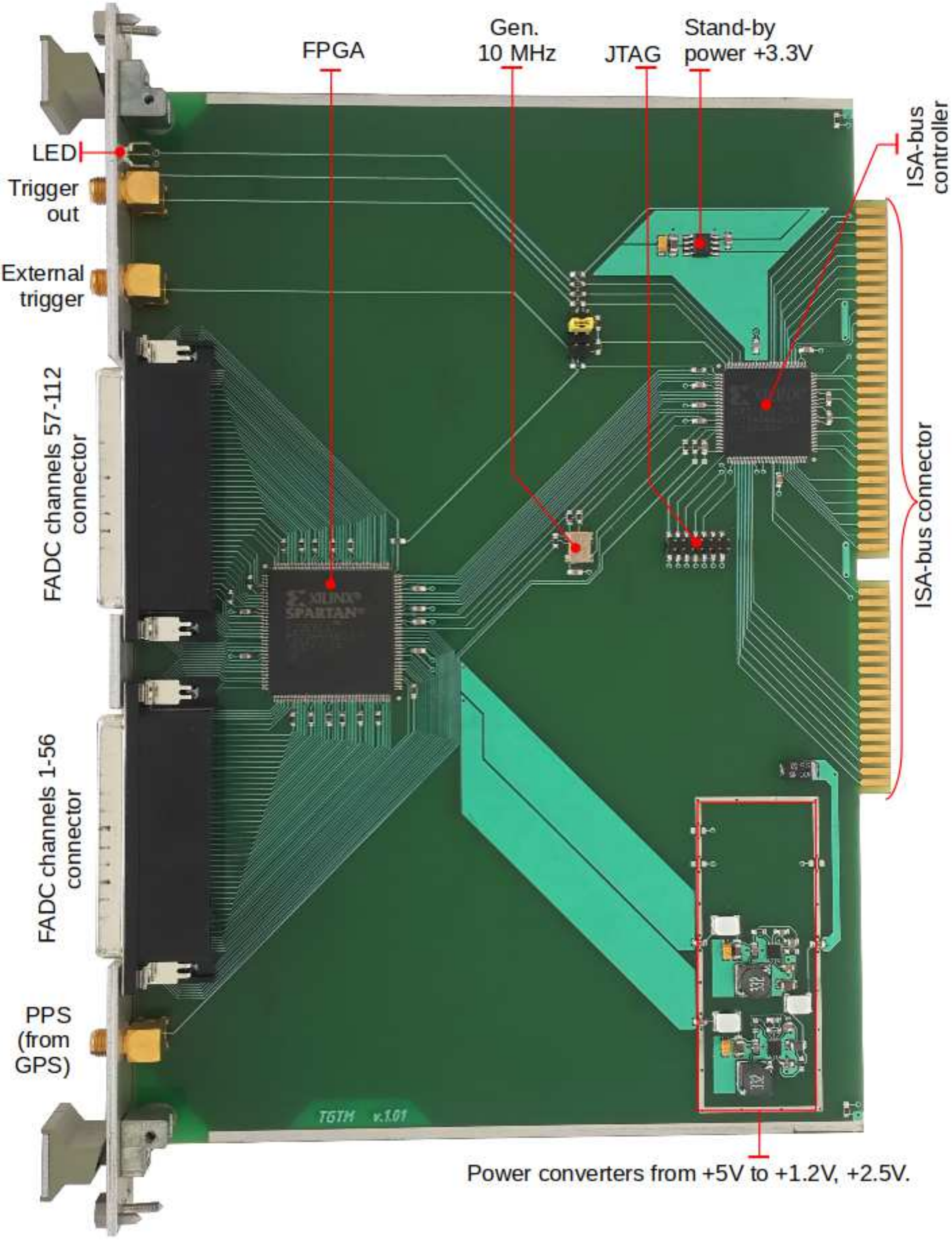}}
\caption{Trigger board.}
\label{trigger_board}
\end{figure}

\subsubsection{Trigger}

The trigger system had 112 input channels (109 main channels and 3 reserved ones) that received discriminator signals from the measuring channels. The trigger board is shown in Fig.~\ref{trigger_board}. The PMT mosaic logical model was uploaded to the XILINX Spartan XC3S400 FPGA chip~\cite{XILINX_Spartan} at the detector boot-up stage. The detector trigger system supported several trigger conditions, some of which could have been active simultaneously. Two ``local'' conditions L2 and L3 (requiring that at least two or three (respectively) adjacent channels to produce discriminator signals within the time window of 1~$\mu$s) and four ``global'' conditions G3, G4, G5, and G7 (requiring that at least three, four, five or seven (respectively) channels with any numbers to produce discriminator signals within the time window of 1~$\mu$s) were available. The trigger system used one local and one global condition simultaneously.

The trigger board had a synchronization port for a GPS module PPS signal input connector. An external trigger input connector allowed to produce forced internal trigger signals that could be be used to record data frames with electronic noise and background light. Additionally, the trigger could have been forced with a software command. The trigger output port was used to send signals to the calibration board to produce calibration frames~\cite{Antonov2016}.

\begin{figure}[t]
\center{\includegraphics[width=0.494\textwidth]{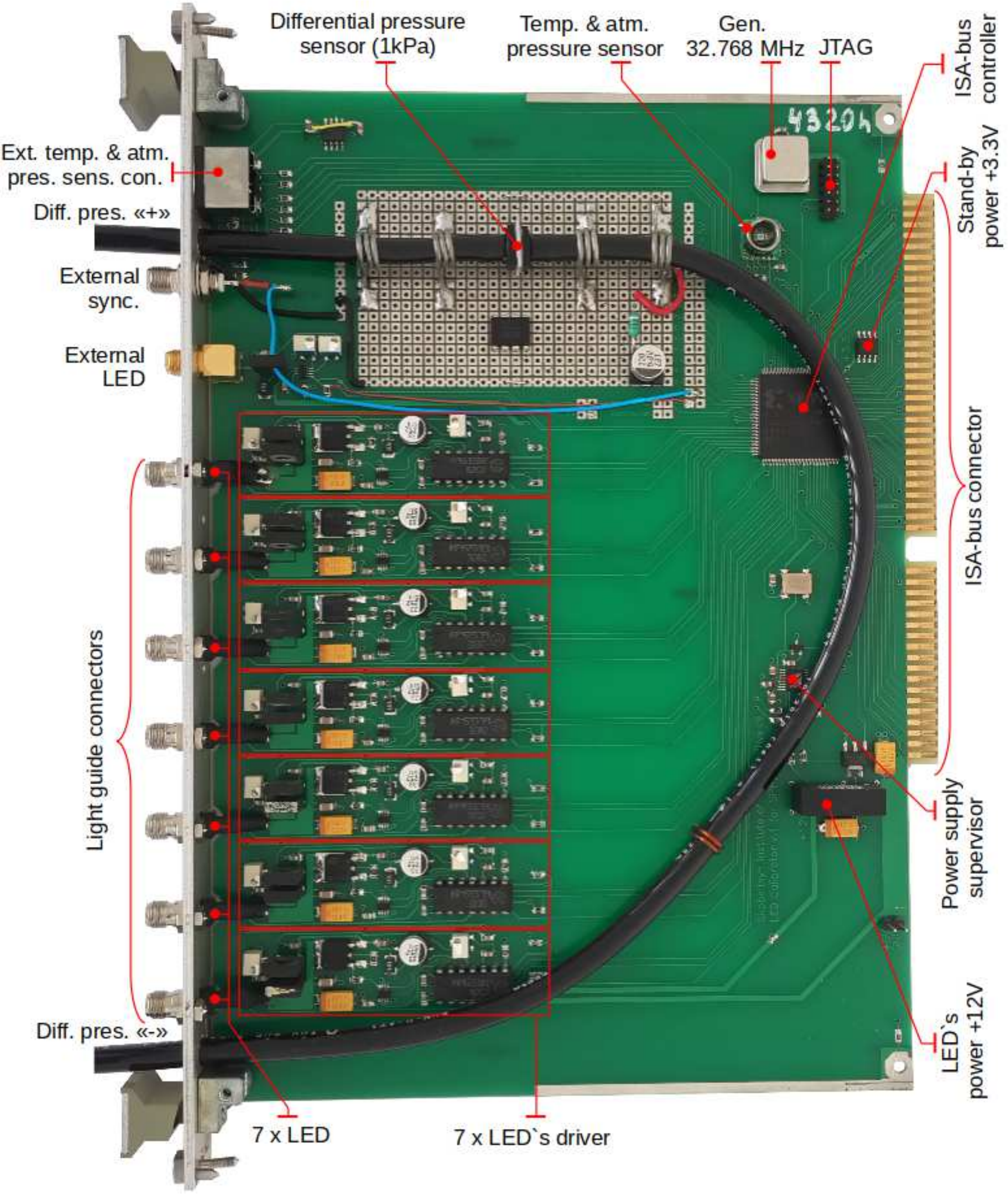}}
\caption{Calibration board.}
\label{led_calibrator_board}
\end{figure}

\subsubsection{Calibration board}
The LED calibration board is an important part of the SPHERE\=/2 detector. This board allowed to obtain an automated on-line relative calibration of each PMT in the mosaic for every detected event. This in turn allowed to reduce the uncertainties in the LDF reconstruction. A detailed description of the SPHERE\=/2 calibration procedures is given in~\cite{Antonov2016}.

The calibration board (shown in Fig.~\ref{led_calibrator_board}) includes seven drivers for Foryard FYL-5013VC1V LEDs~\cite{foryard_cite} with the maximum emittance around 402--405~nm. The LEDs were installed on the board (in the thermostabilized box) and the light was transferred to the mosaic via optical fibers. The LED drivers allowed to form light pulses with a programmed intensity and a nearly rectangular shape. The pulse voltage amplitude was set individually for each LED in the 3.3--8.9~V range using a AD5245BRJ50 DAC~\cite{AD5245BRJ50} (same as in the HVPSs) and voltage stabilizer LM217~\cite{LM217}. The rectangular pulse shape was formed using a differential cascade on KT972A~\cite{Integral_972} and KT973A~\cite{Integral_973} transistors and an `on-off' switch based on a KP1533LE1 chip (an analogue to SN74ALS02~\cite{TI_SN74ALS02A}). The pulse durations and switching order (see table in~\cite{Antonov2016}) were controlled by a calibration board firmware uploaded into a XILINX XCR3128XL CPLD chip~\cite{XCR3128XL}.

The calibration flash series were triggered by an external input from the trigger board or by a software command (used for testing purposes only since this way is relatively slow). The delay between the arrival of the trigger signal and the first LED flash was set in an FPGA firmware. The LED drivers were powered from a +12~V line using a TMA0512S~\cite{TRACO} DC-DC converter. The 8$^{th}$ calibration board channel was reserved for an incandescent lamp used to emulate mosaic illumination by the starlight background. 

A local pressure and temperature sensor HP03S~\cite{HP03S} was installed on the calibration board. Since the conditions in the thermostatic box could differ from the outside conditions, an additional (external) pressure and temperature sensor was utilized, and an additional connector for this sensor was added to the calibration board front panel. As well, a differential pressure sensor CPCL04DFC~\cite{CPCL04DFC} was utilized to control the overpressure inside the balloon (see Section~\ref{sec:aerostat}) during the initial balloon infilling and on every stage of flight. The overpressure rapidly increases with altitude but slowly decreases over time at a constant altitude due to the cooling of the gas in the balloon. At the 750~Pa overpressure, it will be remembered, the safety vent opens, resulting in a loss of gas. Therefore, the overpressure measurements proved to be especially useful during the initial ascent, allowing us to choose such an ascent speed when the gas cools sufficiently to avoid an unnecessary leakage.

\subsubsection{On-board computer}

An Advantech PCA-6781VE~\cite{Advantech} industrial computer with an Intel Celeron~M 600~MHz processor operating under the Slackware Linux 13.37 OS~\cite{Slack} was used as the on-board computer. The computer was directly connected to the measuring boards, trigger and calibration board through an ISA-bus (see Section~\ref{control block}). The computer board front panel housed a power switch, a heater switch, four USB 2.0 ports for a D-Link DWA-126~\cite{D-link_DWA126} Wi-Fi module with flash stick connections and external antenna ANT24-1201~\cite{D-link_ANT24}. Additionally, a keyboard and monitor could have been connected for direct access and OS configuration.

Secondary power supply sources of the detector were located inside the on-board computer module. These power sources were assembled using XP Power~\cite{XP-Power} ICH10024S05 (+5~V, 20~A), JTA1024S12 (+12~V, 0.83~A) and ICH5024WS15 (+15~V, 3,33~A) DC-DC converters on a separate low voltage power source (LVPS) board with a XILINX~\cite{Xilinx} XCR3128XL CPLD chip as a controller. This controller served as a bridge between the LPT port and the I$^2$C interface. The I$^2$C interface emulated on the on-board computer LPT port pins was used to access and control the PMT mosaic commutator (see Section~\ref{sec:commutator}) and cooling system fans, and for measurement of the main power supply batteries voltage and current.

\begin{figure}[t]
\center{\includegraphics[width=0.45\textwidth]{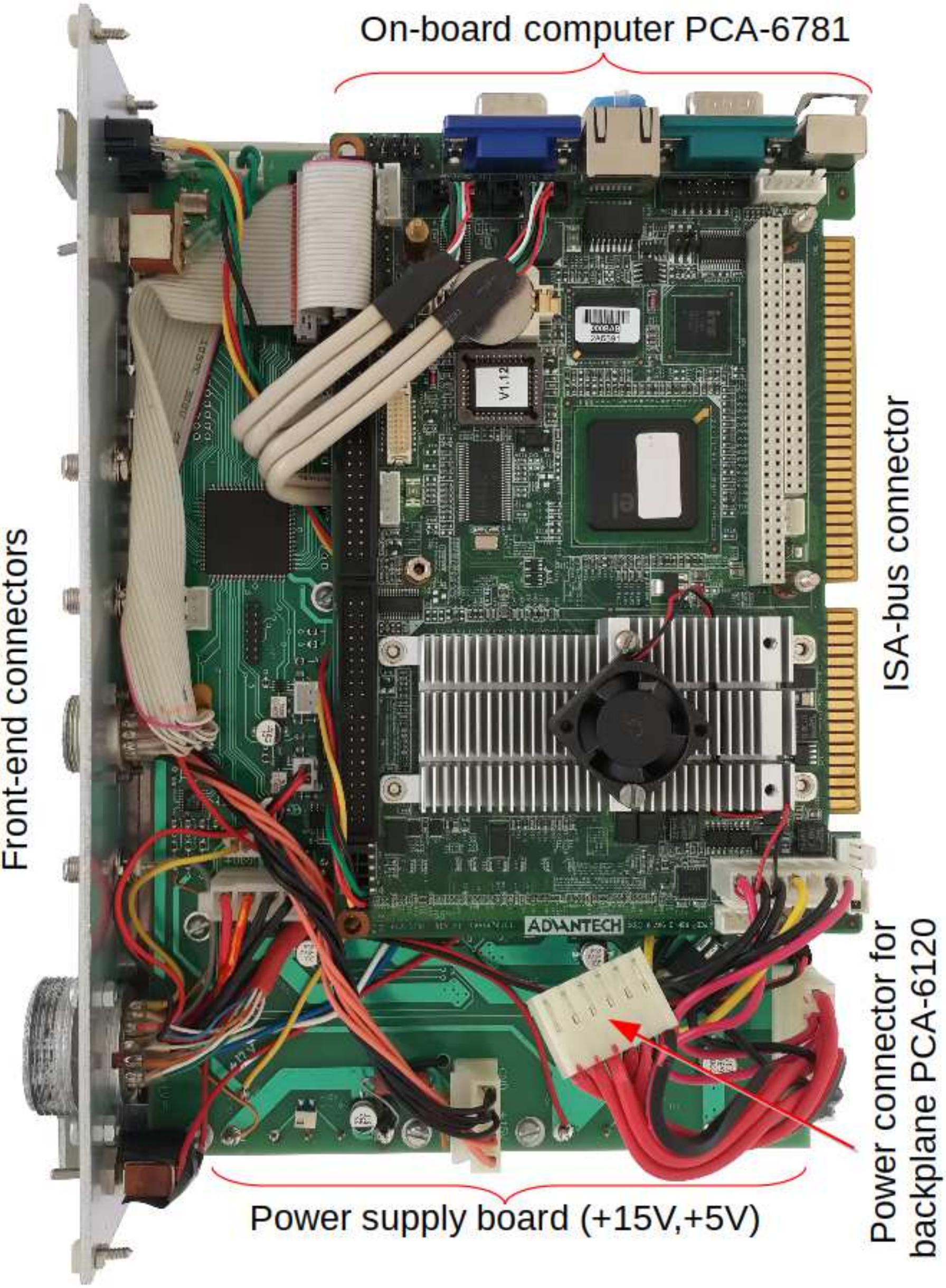}}
\caption{On-board computer and power supply.}
\label{on-board_computer}
\end{figure}

The on-board computer itself was powered from two power sources: +5~V and +12~V (see Fig.~\ref{on-board_computer}). Since the computer operational temperatures were from 0$^\circ$ to 60$^\circ$~C a 25~W electric heater was installed on the back of the computer board to allow normal boot-up at ambient temperatures of about $-20^\circ$~C. Heating required 10-15 minutes and after a successful system boot the heater was switched off since on-board electronics when operating generated enough heat to sustain the required range of temperatures.

\subsection{Data transmission and telemetry}
The SPHERE-2 apparatus had numerous temperature and pressure sensors to control both the flight conditions and the state of electronics, and was designed to be fully automatic. However, the control block also included a Wi-Fi USB-adapter and antenna that allowed to operate the detector remotely from the ground. To maintain communication at distances more than 300~m a directional antenna was employed, therefore the Wi-Fi connection was unstable due to the changes in the position of the SPHERE-2 detector due to the wind. The signal was lost when the detector left the 45$^{\circ}$ cone relative to the antenna. This problem was solved by changing the direction of the antenna manually.

The Wi-Fi link was used to transfer telemetry data, registered event notifications with brief overviews of these events, and also to send commands to the control block (including trigger state updates, PMTs powering, trigger conditions, etc.). The altitude and coordinates of the SPHERE-2 detector were controlled by a Garmin 16HVS GPS~\cite{GarminGPS} through a serial RS-232 port. The GPS had a time PPS output with the 1~$\mu{}s$ precision to link detected events to the world time. 
The HP03S pressure sensor~\cite{HP03S} the measured atmospheric pressure outside the block. A differential pressure sensor CPCL04DFC~\cite{CPCL04DFC} with instrumentation amplifier INA128P~\cite{INA128P} allowed to measure the overpressure inside the balloon shell with the accuracy better than 3~Pa. The PMT mosaic was supplemented with its own Analog Devices AD7415~\cite{AD7415} temperature sensor.

\subsection{Power supply system}

Behind the electronics six UNB-01 (EL1901) Li-Ion battery packs with a total capacity of about 900~W/h were installed in a separate volume. The power supply system of the SPHERE-2 detector used several batteries in the pack and was designed to allow ``hot'' exchange of the batteries in the pack without the need to turn off the detector. At maximum power consumption by the apparatus being about 50~W the power system guaranteed 15~hours of stable work. At minimum power consumption (on-board computer active, communication systems active, temperature control active, all other systems disabled) the batteries could power the detector for about 50~hours. The second pack of six batteries and the hot-reload capability of the power system allowed to keep the apparatus on-line during the full duration of the measurement run (i.e. about 10 days).

\section{Discussion and conclusions}\label{sec:concl}

In this paper we provide a description of the SPHERE experiment, accounting for all changes in the SPHERE-2 detector design that were made after test flights of 2008-2010. We performed a detailed simulation of the SPHERE-2 detector response, including a calculation of its instrumental acceptance (see~\cite{ANTONOV201924}). Three successful measurement runs were carried out in 2011-2013 at Lake Baikal, resulting in more than 10$^{3}$ registered EAS events~\cite{Antonov:2015xta}. Analysis and interpretation of experimental data is provided in~\cite{Antonov:2015xta,Antonov:2015aqa}. In these papers it was shown that the method of EAS registration using reflected Cherenkov light can be used not only to reconstruct the energy of the primary particle, but also to assess the CR composition. A more detailed analysis with refined account of systematics is in progress; its results will be published elsewhere.

Unfortunately, the total number of showers registered with the SPHERE-2 detector is still rather small compared to ground-based arrays such as KASCADE-Grande~\cite{ape12}, EAS-TOP~\cite{agl89} and TALE~\cite{abb18}. This is mainly due to two reasons, namely: 1) modest sensitive area and quantum efficiency of the FEU-84-3 PMT used in the experiment, 2) technical difficulties with maintaining the ``BAPA'' balloon in operational conditions and some logistic issues. The use of more sensitive PMTs would increase the number of recorded events several times. Furthermore, the conditions of the Russian winter make allow up to five measurement runs per year, given an adequate supply of helium. In total, the expected number of events after these improvements could rise by about an order of magnitude.

The development of the SPHERE-2 detector hardware was, for the most part, carried out in the period of 2005-2010 with the use of existing and available at that time electronic components and materials. Currently, with the increase in functionality and miniaturization of electronics, as well as with the advent of new silicon photo multipliers, it is possible to achieve the same technical goals in simpler and more effective ways. The use of silicon PMT and modern electronic components would reduce the weight of the equipment by at least 10 times. In turn, the smaller mass of the detector would allow to abandon the cumbersome balloon equipment and switch to the use of unmanned aerial vehicles such as a copter or a drone as a carrier.

\section{Acknowledgments}
We are grateful to group of S.B.~Shaulov from the Lebedev Physical Institute of the Russian Academy of Sciences for their assistance in assembling and testing of the electronic equipment and preparation of the expeditions. We also thank the Baikal-GVD collaboration and G.V.~Domogatsky (Institute for Nuclear Research, Russian Academy of Sciences) for the support of the SPHERE experiment at the Baikal Lake scientific station. This work was supported by the Russian Foundation for Basic Research [grants 11-02-01475-a, 12-02-10015-k] and the Presidium of the Russian Academy of Sciences. The work of T.D. was supported by the Munich Institute for Astro- and Particle Physics (MIAPP) of the DFG cluster of excellence ``Origin and Structure of the Universe''.

\bibliographystyle{elsarticle-num}
\bibliography{Sphere-Detector}

\begin{thebibliography}{10}
\expandafter\ifx\csname url\endcsname\relax
  \def\url#1{\texttt{#1}}\fi
\expandafter\ifx\csname urlprefix\endcsname\relax\def\urlprefix{URL }\fi
\expandafter\ifx\csname href\endcsname\relax
  \def\href#1#2{#2} \def\path#1{#1}\fi

\bibitem{Antoni2005}
T.~Antoni, et~al. KASCADE~collaboration, {KASCADE} measurements of energy
  spectra for elemental groups of cosmic rays: Results and open problems,
  Astroparticle Physics 24~(1-2) (2005) 1--25.
\newblock \href {http://dx.doi.org/10.1016/j.astropartphys.2005.04.001}
  {\path{doi:10.1016/j.astropartphys.2005.04.001}}.

\bibitem{Aglietta2004}
M.~Aglietta, et~al., The cosmic ray primary composition in the
  {\textquotedblleft}knee{\textquotedblright} region through the {EAS}
  electromagnetic and muon measurements at {EAS}-{TOP}, Astroparticle Physics
  21~(6) (2004) 583--596.
\newblock \href {http://dx.doi.org/10.1016/j.astropartphys.2004.04.005}
  {\path{doi:10.1016/j.astropartphys.2004.04.005}}.

\bibitem{Fowler2001}
J.~Fowler, et~al., A measurement of the cosmic ray spectrum and composition at
  the knee, Astroparticle Physics 15~(1) (2001) 49--64.
\newblock \href {http://dx.doi.org/10.1016/s0927-6505(00)00139-0}
  {\path{doi:10.1016/s0927-6505(00)00139-0}}.

\bibitem{ANTONOV201924}
R.~Antonov, et~al., {Spatial and temporal structure of EAS reflected Cherenkov
  light signal}, Astroparticle Physics 108 (2019) 24 -- 39.
\newblock \href {http://dx.doi.org/10.1016/j.astropartphys.2019.01.002}
  {\path{doi:10.1016/j.astropartphys.2019.01.002}}.

\bibitem{Antonov:2015xta}
R.~A. Antonov, et~al., {Detection of reflected Cherenkov light from extensive
  air showers in the SPHERE experiment as a method of studying superhigh energy
  cosmic rays}, Phys. Part. Nucl. 46~(1) (2015) 60--93.
\newblock \href {http://dx.doi.org/10.1134/S1063779615010025}
  {\path{doi:10.1134/S1063779615010025}}.

\bibitem{Antonov:2015aqa}
R.~A. Antonov, et~al., {Event-by-event study of CR composition with the SPHERE
  experiment using the 2013 data}, J. Phys. Conf. Ser. 632~(1) (2015) 012090.
\newblock \href {http://arxiv.org/abs/1503.04998} {\path{arXiv:1503.04998}},
  \href {http://dx.doi.org/10.1088/1742-6596/632/1/012090}
  {\path{doi:10.1088/1742-6596/632/1/012090}}.

\bibitem{sphere2009FIAN-eng}
A.~Anokhina, et~al., {Method for measuring the PCR proton spectrum in the
  energy range of $ > 10^{16}$ eV}, Bulletin of the Lebedev Physics Institute
  36~(5) (2009) 146--149.
\newblock \href {http://dx.doi.org/10.3103/s1068335609050042}
  {\path{doi:10.3103/s1068335609050042}}.

\bibitem{sphere2013JP-results}
R.~Antonov, et~al., {Results on the primary CR spectrum and composition
  reconstructed with the SPHERE-2 detector}, Journal of Physics CS 409~(1)
  (2013) 012088--012091.
\newblock \href {http://dx.doi.org/10.1088/1742-6596/409/1/012088}
  {\path{doi:10.1088/1742-6596/409/1/012088}}.

\bibitem{GVD2018}
G.~Domogatsky, et~al., {Gigaton Volume Detector in Lake Baikal: status of the
  project}, 2018, p. 063.
\newblock \href {http://dx.doi.org/10.22323/1.307.0063}
  {\path{doi:10.22323/1.307.0063}}.

\bibitem{rosaerosystems}
{Augur RosAeroSystems}, \href{http://rosaerosystems.com}{{Website}}.
\newline\urlprefix\url{http://rosaerosystems.com}

\bibitem{lamcotec}
L.~Inc., \href{http://www.lamcotec.com/}{Laminating coating technologies inc,
  website}.
\newline\urlprefix\url{http://www.lamcotec.com/}

\bibitem{I2C_spec}
\href{http://www.i2c-bus.org/}{{“Interface specification.”}}.
\newline\urlprefix\url{http://www.i2c-bus.org/}

\bibitem{OSLO}
{Optics Software for Layout and Optimization},
  \href{https://www.lambdares.com/oslo}{{Website}}.
\newline\urlprefix\url{https://www.lambdares.com/oslo}

\bibitem{FEU84}
P.~Dunaevskaya, M.~Podoksina, J.~Ronkin, {Photoelectron multiplier {FEU-84}},
  Pribory i tehnika experimenta (in Russian). 5 (1970) 252--255.

\bibitem{R3886}
{Hamamatsu Photonics K.K.},
  \href{http://www.hamamatsu.com/resources/pdf/etd/L11416\_L11494\_TACC1057E.pdf}{{Product
  specification.}} (2010).
\newline\urlprefix\url{http://www.hamamatsu.com/resources/pdf/etd/L11416\_L11494\_TACC1057E.pdf}

\bibitem{Antonov2016}
R.~Antonov, et~al., {The LED calibration system of the SPHERE-2 detector},
  Astroparticle Physics 77 (2016) 55–65.
\newblock \href {http://dx.doi.org/10.1016/j.astropartphys.2016.01.004}
  {\path{doi:10.1016/j.astropartphys.2016.01.004}}.

\bibitem{murata}
{Murata},
  \href{https://www.murata.com/~/media/webrenewal/support/library/catalog/products/capacitor/mlcc/
  c02e.ashx?la=en-us}{{“Product specification”}}.
\newline\urlprefix\url{https://www.murata.com/~/media/webrenewal/support/library/catalog/products/capacitor/mlcc/
  c02e.ashx?la=en-us}

\bibitem{MCC}
{Micro Commercial Components Corp.},
  \href{https://www.mccsemi.com/pdf/Products/SM4001PL-SM4007PL(SOD-123FL)-V3.pdf}{{“Product
  specification”}}.
\newline\urlprefix\url{https://www.mccsemi.com/pdf/Products/SM4001PL-SM4007PL(SOD-123FL)-V3.pdf}

\bibitem{MAX1847}
{Maxim Integrated},
  \href{https://datasheets.maximintegrated.com/en/ds/MAX1846-MAX1847.pdf}{{“Product
  specification”}}.
\newline\urlprefix\url{https://datasheets.maximintegrated.com/en/ds/MAX1846-MAX1847.pdf}

\bibitem{AD5245BRJ50}
{Analog Devices},
  \href{http://www.analog.com/media/en/technical-documentation/data-sheets/AD5245.pdf}{{“Product
  specification”}}.
\newline\urlprefix\url{http://www.analog.com/media/en/technical-documentation/data-sheets/AD5245.pdf}

\bibitem{AD7994BRU}
{Analog Devices},
  \href{http://www.analog.com/media/en/technical-documentation/data-sheets/AD7993\_7994.pdf}{{“Product
  specification”}}.
\newline\urlprefix\url{http://www.analog.com/media/en/technical-documentation/data-sheets/AD7993\_7994.pdf}

\bibitem{EPCOS}
{EPCOS AG},
  \href{http://datasheetz.com/data/Sensors,%20Transducers/Thermistors%20-%20NTC/B57621C474J62-datasheetz.html}{{“Product
  specification (not in production now)”}}.
\newline\urlprefix\url{http://datasheetz.com/data/Sensors,%20Transducers/Thermistors%20-%20NTC/B57621C474J62-datasheetz.html}

\bibitem{ADuM1251}
{Analog Devices},
  \href{http://www.analog.com/media/en/technical-documentation/data-sheets/ADUM1250\_1251.pdf}{{“Product
  specification”}}.
\newline\urlprefix\url{http://www.analog.com/media/en/technical-documentation/data-sheets/ADUM1250\_1251.pdf}

\bibitem{PCA9547}
{NXP Semiconductors},
  \href{https://www.nxp.com/docs/en/data-sheet/PCA9547.pdf}{{“Product
  specification”}}.
\newline\urlprefix\url{https://www.nxp.com/docs/en/data-sheet/PCA9547.pdf}

\bibitem{MF-R005}
{Bourns, Inc.},
  \href{http://bourns.com/docs/product-datasheets/mfr.pdf?sfvrsn=bc732717\_30}{{“Product
  specification”}}.
\newline\urlprefix\url{http://bourns.com/docs/product-datasheets/mfr.pdf?sfvrsn=bc732717\_30}

\bibitem{AD7415}
{Analog Devices},
  \href{http://www.analog.com/media/en/technical-documentation/data-sheets/AD7414\_7415.pdf}{{“Product
  specification”}}.
\newline\urlprefix\url{http://www.analog.com/media/en/technical-documentation/data-sheets/AD7414\_7415.pdf}

\bibitem{HMC6352}
{Honeywell},
  \href{https://datasheet4u.com/datasheet-pdf/Honeywell/HMC6352/pdf.php?id=632821}{{HMC6352
  Datasheet}}.
\newline\urlprefix\url{https://datasheet4u.com/datasheet-pdf/Honeywell/HMC6352/pdf.php?id=632821}

\bibitem{inclinometer}
{Durham Instruments},
  \href{https://disensors.com/product/dql-dual-axis-inclinometer/}{{“Product
  specification”}}.
\newline\urlprefix\url{https://disensors.com/product/dql-dual-axis-inclinometer/}

\bibitem{Advantech}
{Advantech Co., Ltd.}, \href{http://www.advantech.com/}{Website}.
\newline\urlprefix\url{http://www.advantech.com/}

\bibitem{AD9203ARU}
{Analog Devices},
  \href{http://www.analog.com/media/en/technical-documentation/data-sheets/AD9203.pdf}{{“Product
  specification”}}.
\newline\urlprefix\url{http://www.analog.com/media/en/technical-documentation/data-sheets/AD9203.pdf}

\bibitem{AD8011}
{Analog Devices},
  \href{http://www.analog.com/media/en/technical-documentation/data-sheets/AD8011.pdf}{{“Product
  specification”}}.
\newline\urlprefix\url{http://www.analog.com/media/en/technical-documentation/data-sheets/AD8011.pdf}

\bibitem{Xilinx}
{Xilinx Inc.}, \href{https://www.xilinx.com/}{{Website}}.
\newline\urlprefix\url{https://www.xilinx.com/}

\bibitem{XILINX_Spartan}
{Xilinx},
  \href{https://www.xilinx.com/support/documentation/data\_sheets/ds099.pdf}{{“Product
  specification”}}.
\newline\urlprefix\url{https://www.xilinx.com/support/documentation/data\_sheets/ds099.pdf}

\bibitem{XCR3128XL}
{Xilinx},
  \href{https://www.xilinx.com/support/documentation/data\_sheets/ds016.pdf}{{“Product
  specification”}}.
\newline\urlprefix\url{https://www.xilinx.com/support/documentation/data\_sheets/ds016.pdf}

\bibitem{MAX1556}
{Maxim Integrated},
  \href{https://datasheets.maximintegrated.com/en/ds/MAX1556-MAX1557.pdf}{{“Product
  specification”}}.
\newline\urlprefix\url{https://datasheets.maximintegrated.com/en/ds/MAX1556-MAX1557.pdf}

\bibitem{ADM8660}
{Analog Devices},
  \href{https://www.analog.com/media/en/technical-documentation/data-sheets/ADM660\_8660.pdf}{{“Product
  specification”}}.
\newline\urlprefix\url{https://www.analog.com/media/en/technical-documentation/data-sheets/ADM660\_8660.pdf}

\bibitem{foryard_cite}
{Ningbo Foryard Optoelectronics},
  \href{{https://datasheet4u.com/datasheet-pdf/NingboForyard/FYL-5013VC1C/pdf.php?id=1096448}}{{FYL-5013VC1C
  Datasheet}}.
\newline\urlprefix\url{{https://datasheet4u.com/datasheet-pdf/NingboForyard/FYL-5013VC1C/pdf.php?id=1096448}}

\bibitem{LM217}
{STMicroelectronics},
  \href{www.st.com/resource/en/datasheet/lm217.pdf}{{“Product
  specification”}}.
\newline\urlprefix\url{www.st.com/resource/en/datasheet/lm217.pdf}

\bibitem{Integral_972}
{JSC INTEGRAL},
  \href{https://www.integral.by/sites/default/files/pdf/kt972.pdf}{{“Product
  specification (in Russian)”}}.
\newline\urlprefix\url{https://www.integral.by/sites/default/files/pdf/kt972.pdf}

\bibitem{Integral_973}
{JSC INTEGRAL},
  \href{https://www.integral.by/sites/default/files/pdf/kt973.pdf}{{“Product
  specification (in Russian)”}}.
\newline\urlprefix\url{https://www.integral.by/sites/default/files/pdf/kt973.pdf}

\bibitem{TI_SN74ALS02A}
{Texas Instruments}, \href{https://www.ti.com/lit/gpn/SN74ALS02A}{{“Product
  specification”}}.
\newline\urlprefix\url{https://www.ti.com/lit/gpn/SN74ALS02A}

\bibitem{TRACO}
{TRACO Electronic AG},
  \href{https://www.tracopower.com/products/tma.pdf}{{“Product
  specification”}}.
\newline\urlprefix\url{https://www.tracopower.com/products/tma.pdf}

\bibitem{HP03S}
{HOPE Microelectronics CO.},
  \href{https://datasheet4u.com/datasheet-pdf/HOPERF/HP03S/pdf.php?id=748092}{{HP03S
  Datasheet}}.
\newline\urlprefix\url{https://datasheet4u.com/datasheet-pdf/HOPERF/HP03S/pdf.php?id=748092}

\bibitem{CPCL04DFC}
{Honeywell International, Inc.},
  \href{https://www.elfa.se/Web/Downloads/\_t/ds/CPC\_eng\_tds.pdf}{{“Product
  specification. Not in production now.”}}.
\newline\urlprefix\url{https://www.elfa.se/Web/Downloads/\_t/ds/CPC\_eng\_tds.pdf}

\bibitem{Slack}
{Slackware Linux, Inc.}, \href{http://www.slackware.com/}{{“OS distribution
  developer”}}.
\newline\urlprefix\url{http://www.slackware.com/}

\bibitem{D-link_DWA126}
{D-Link (Europe) Ltd.},
  \href{http://dlink.ru/mn/products/2/1297.html}{{“Product
  specification”}}.
\newline\urlprefix\url{http://dlink.ru/mn/products/2/1297.html}

\bibitem{D-link_ANT24}
{D-Link (Europe) Ltd.},
  \href{http://www.dlink.ru/mn/products/2/229\_b.html}{{“Product
  specification”}}.
\newline\urlprefix\url{http://www.dlink.ru/mn/products/2/229\_b.html}

\bibitem{XP-Power}
{XP Power}, \href{https://www.xppower.com/}{{Website}}.
\newline\urlprefix\url{https://www.xppower.com/}

\bibitem{GarminGPS}
{Garmin Ltd.},
  \href{http://static.garmin.com/pumac/470\_GPS16\_17TechnicalSpecification.pdf}{{“Product
  specification”}}.
\newline\urlprefix\url{http://static.garmin.com/pumac/470\_GPS16\_17TechnicalSpecification.pdf}

\bibitem{INA128P}
{Texas Instruments}, \href{https://www.ti.com/lit/gpn/INA128}{{“Product
  specification”}}.
\newline\urlprefix\url{https://www.ti.com/lit/gpn/INA128}

\bibitem{ape12}
W.~Apel, et~al. (The KASCADE-Grande~Collaboration), {The spectrum of
  high-energy cosmic rays measured with KASCADE-Grande}, APh 36 (2012)
  183--194.
\newblock \href {http://dx.doi.org/10.1016/j.astropartphys.2012.05.023}
  {\path{doi:10.1016/j.astropartphys.2012.05.023}}.

\bibitem{agl89}
{M. Aglietta et al. (The EAS-TOP Collaboration)}, {The EAS-TOP array at
  E$_{0}$= 10$^{14}$--10$^{16}$ eV: Stability and Resolutions}, NIM A 277
  (1989) 23--28.
\newblock \href {http://dx.doi.org/10.1016/0168-9002(89)90531-7}
  {\path{doi:10.1016/0168-9002(89)90531-7}}.

\bibitem{abb18}
R.~Abbasi, et~al. The TALE~Collaboration, {The Cosmic-Ray Energy Spectrum
  between 2 PeV and 2 EeV Observed with the TALE detector in monocular mode},
  ApJ 865 (2018) 74.
\newblock \href {http://dx.doi.org/10.3847/1538-4357/aada05}
  {\path{doi:10.3847/1538-4357/aada05}}.

\end{thebibliography}
\end{document}